# Microscopic crystallographic analysis of dislocations in molecular crystals

Sang T. Pham[1], Natalia Koniuch[1], Emily Wynne[1],
Andy Brown[1], Sean M. Collins[1,2]*

[1]*Bragg Centre for Materials Research & School of Chemical and Process Engineering, University of Leeds, Woodhouse Lane, Leeds LS2 9JT, UK*
[2]*School of Chemistry, University of Leeds, Woodhouse Lane, Leeds LS2 9JT, UK*

*Email: s.m.collins@leeds.ac.uk

**Organic molecular crystals encompass a vast range of materials from pharmaceuticals to organic optoelectronics and proteins to waxes in biological and industrial settings. Crystal defects from grain boundaries to dislocations are known to play key roles in mechanisms of growth[1,2] and also in the functional properties of molecular crystals[3–5]. In contrast to the precise analysis of individual defects in metals, ceramics, and inorganic semiconductors enabled by electron microscopy, significantly greater ambiguity remains in the experimental determination of individual dislocation character and slip systems in molecular materials[3]. In large part, nanoscale dislocation analysis in molecular crystals has been hindered by the severely constrained electron exposures required to avoid irreversibly degrading these crystals[6]. Here, we present a low-dose, single-exposure approach enabling nanometre-resolved analysis of individual extended dislocations in molecular crystals. We demonstrate the approach for a range of crystal types to reveal dislocation character and operative slip systems unambiguously.**

Across materials classes, dislocations in crystals have a profound effect on functional properties. Mechanical properties of metals are intimately linked with dislocation motion and pinning[7,8]. While typically seen as deleterious to performance in semiconductors, such as in reduced efficiencies in GaN-based light emitting diodes[9], they have been identified as a means for advantageous modulation of ionic charge transport properties in oxides[10]. These findings have relied heavily on precise imaging and diffraction studies by electron microscopy for analysis of individual dislocations. While scanning probe microscopy,[11] X-ray topography[12], and modelling[3] techniques have supported dislocation characterisation in organic molecular crystals, these techniques lack the spatial resolution to image individual dislocations at high density.

Dislocations are in-effect a planar cut through a crystal followed by atomic displacements in a particular direction, the Burgers vector **B** (Fig. 1a). The singular displacement direction creates a linear defect, and this dislocation line vector **u** coincides with the location of maximum local lattice distortion in the crystal. Together, these vectors describe the operative slip systems and the character of the dislocation, i.e. edge, screw, or mixed dislocations. For edge dislocations **B** is perpendicular to **u**, while for screw dislocations **B** is parallel to **u** (Fig. 1a). The classical electron microscopy approach to characterise a dislocation involves the combination of electron diffraction and diffraction contrast imaging to evaluate the behaviour of sets of lattice planes and associated diffraction vectors $\mathbf{g}_{hkl}$ in the vicinity of the dislocation. For $\mathbf{g}_{hkl} \cdot (\mathbf{B} \times \mathbf{u}) = 0$, a unique invisibility criterion is established for $\mathbf{g}_{hkl} \cdot \mathbf{B} = 0$. By exploring a range of diffraction conditions, the Burgers vector direction can be determined unambiguously.

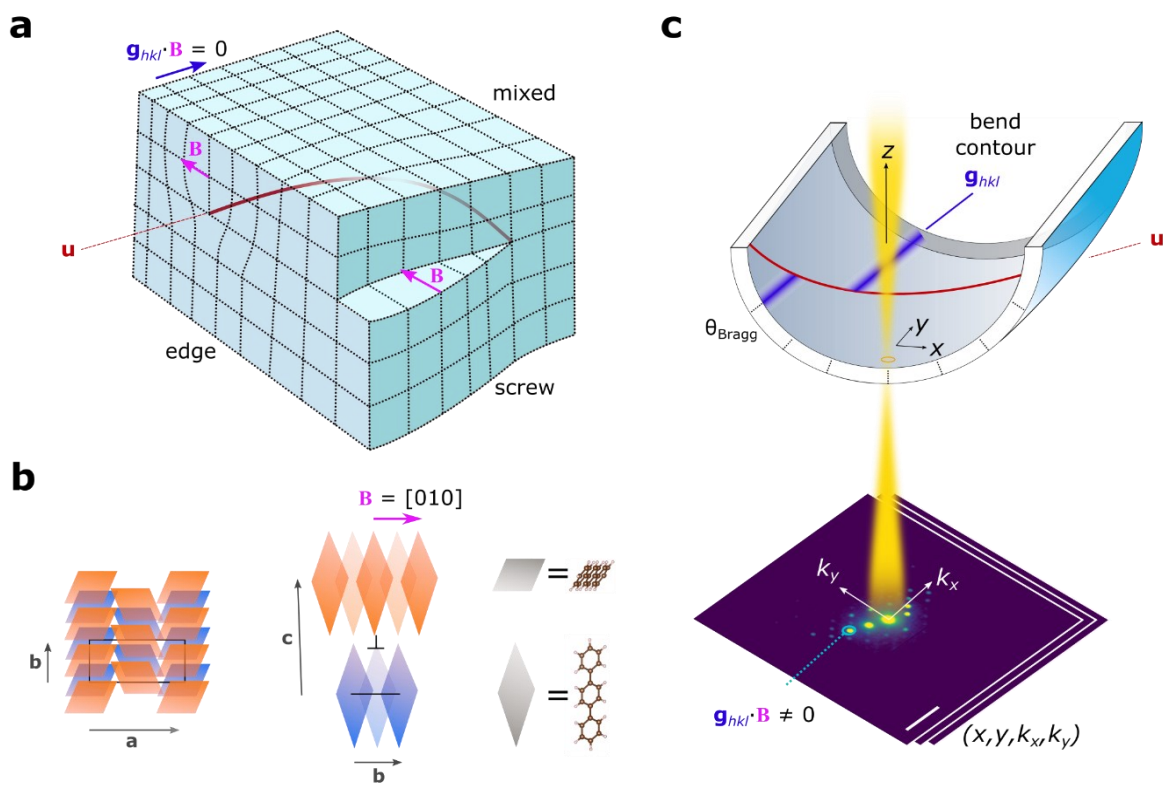

**Figure 1. Low-dose dislocation analysis by scanning electron diffraction. a**, Illustration of the distortion of planes due to edge, mixed, and screw dislocations, with no distortion for planes corresponding to diffraction vectors $\mathbf{g}_{hkl}$ at the invisibility criterion. **b**, Schematic of an edge dislocation core in a p-terphenyl crystal with Burgers vector **B** = [010]. The blue parallelograms represent molecules in a plane below those in orange. A black rectangle or line marks the *ab* plane of the p-terphenyl lattice in the plane below (blue) the dislocation core (⊥). **c**, A bend contour appears where the sample curvature brings planes to the exact Bragg condition, with displacements in the bend contour due to abrupt changes in the orientation of planes around the dislocation core, marked by the dislocation line **u**. The electron beam is scanned in (*x, y*) enabling reconstruction of bend contours from multiple diffraction vectors $\mathbf{g}_{hkl}$ in parallel from two-dimensional diffraction patterns ($k_x, k_y$) recorded at each probe position to give a four-dimensional dataset ($x, y, k_x, k_y$). The sample bending is exaggerated for visual effect.

Atomic resolution imaging of dislocations[13], including at fluences <200 e$^-$ Å$^{-2}$ in halide perovskites[14], as well as three-dimensional imaging modes[15,16] have also been demonstrated. However, none of the existing approaches support dislocation analysis at ~10 e$^-$ Å$^{-2}$ as required for electron microscopy of many organic molecular crystals[6]. Moreover, the complex molecular packing of organic crystals in projection (Fig. 1b) generally precludes lattice imaging of dislocations beyond a few high-symmetry orientations for pure edge dislocations[17,18]. Early work on real space crystallography supported inference of plausible Burgers vector directions in molecular crystals[19–21], but these dislocation analyses have only been reported for organic molecular crystals with relatively high stability under high energy electron beam exposure such as p-terphenyl (Fig. 1b).

Scanning electron diffraction (SED), a type of four-dimensional scanning transmission electron microscopy (4D-STEM) with a nearly parallel, low convergence angle probe,[22] suggests a route to low-dose electron microscopy for dislocation analysis. SED combines real and reciprocal space data-acquisition simultaneously and can be operated at sufficiently low electron fluences to detect mosaicity in peptide crystals[23] and to examine the role of molecular packing in energy transport in organic semiconductors[24–26]. We now establish a methodology for single-exposure SED analysis of extended, in-plane dislocations in thin, electron transparent organic crystals at electron fluences as low as 5 e$^-$ Å$^{-2}$. We apply this approach to aromatic molecules, long-chain hydrocarbons, and hydrogen-bonding crystals as key organic optoelectronic, wax, and pharmaceutical model systems. The methodology enables the determination of the operative slip systems for orientations on and off high symmetry zone axes and the identification of edge, screw, and mixed type dislocation character as well as the handedness of screw components.

SED enables the parallel acquisition of many diffraction vector intensities by recording a two-dimensional diffraction pattern at each probe position in a two-dimensional scan (Fig. 1c). By selecting individual $\mathbf{g}_{hkl}$, we construct virtual dark field (VDF) images from the four-dimensional dataset. For electron-transparent crystals deposited onto electron microscopy grids, bending of the crystals results in the appearance of bands marking the positions in the sample that are exactly at the Bragg condition for a specific $\mathbf{g}_{hkl}$, denoted bend contours. For planes that are distorted due to the presence of a dislocation ($\mathbf{g}_{hkl} \cdot \mathbf{B} \neq 0$) a break in the otherwise continuous bend contour appears at the dislocation line $\mathbf{u}$[19,20,27], akin to features observed in convergent beam electron diffraction of dislocations[28]. Such a break corresponds to a displacement of the bend contour along $\mathbf{u}$ (Supplementary Note 1). For $\mathbf{g}_{hkl}$ at the invisibility criterion, the bend contours remain continuous.

In SED, all recorded $\mathbf{g}_{hkl}$ in the diffraction plane of sufficient signal-to-noise ratio can be used to construct VDF images for this bend contour analysis as a function of the azimuthal angle in the diffraction plane φ (Supplementary Figure 1). A geometric model of the tilt of lattice planes *hkl* across the dislocation core describes the resulting variation in the bend contour displacement *f(φ)* for in-plane dislocations (Supplementary Figure 2):

$$f(\varphi) = A \arctan(B \cos^2(\varphi - C)) \qquad (1)$$

where *A*, *B*, and *C* are fitting coefficients related to the local radius of curvature, the Burgers vector, and the azimuthal orientation of the diffraction pattern on the detector, respectively. The displacement *f(φ)* is zero at the invisibility criterion and maximal along the Burgers vector direction. The crystal orientation ascertained from the electron diffraction data together with the Burgers vector direction (equation 1) provides the necessary information to identify the operative slip system (Supplementary Note 1) as well as the dislocation character by inspection of the relative directions of **B** and **u.**

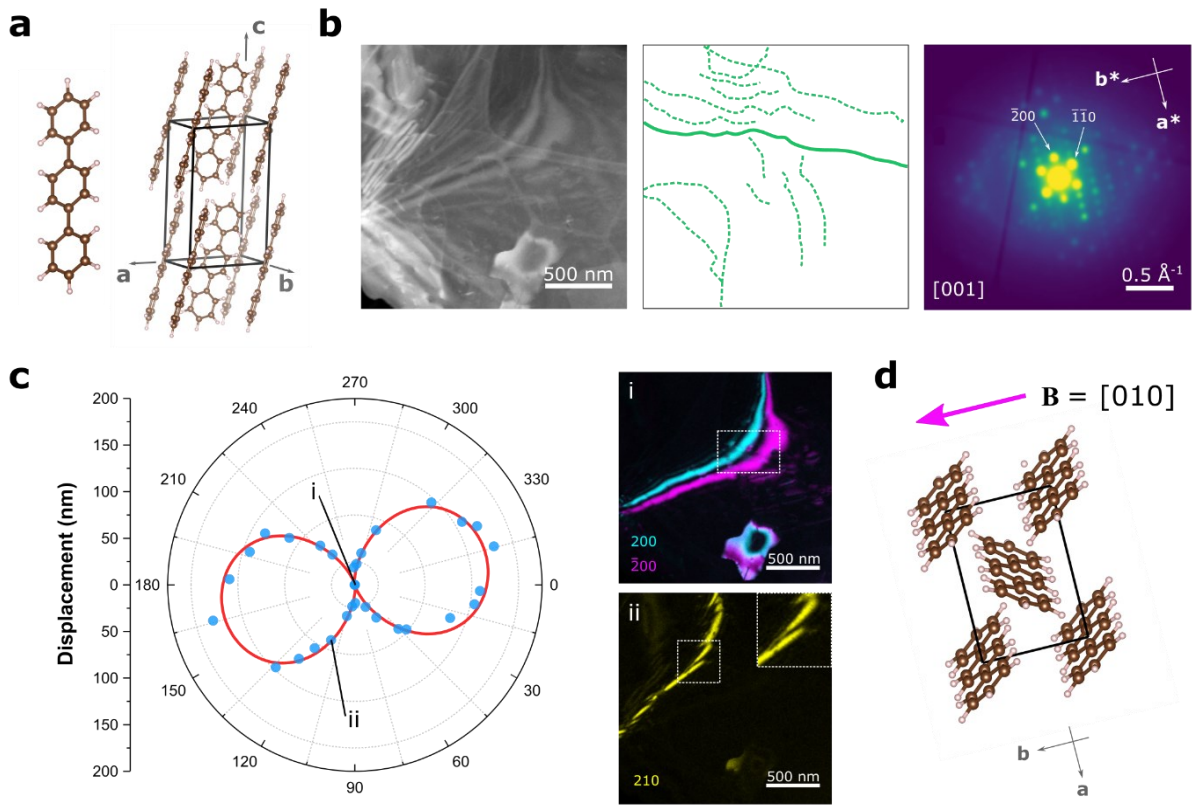

**Figure 2. Dislocation analysis in p-terphenyl. a**, A representation of the p-terphenyl molecule and the p-terphenyl unit cell. **b**, An ADF-STEM image, dislocation network (green lines tracing diffraction contrast features), and average diffraction pattern extracted from a single SED (4D-STEM) dataset. The diffraction pattern indexes to the [001] zone axis. **c**, A polar plot of the bend contour displacements at the solid dislocation line highlighted in **b** for a series of $g_{hkl}$ as a function of their azimuthal angle φ. The red trace marks a fit to equation 1 with the measured displacements in blue. Two VDF images corresponding to the displacements marked (i, ii) depicting bend contours at the $g_{hkl}·\mathbf{B} = 0$ and $g_{hkl}·\mathbf{B} \neq 0$ conditions. The *hkl* are marked on the individual VDF images. **d**, The p-terphenyl unit cell oriented to match the sample orientation. The magenta arrow marks the Burgers vector **B** = [010].

Figure 2 presents a first demonstration of the approach applied to p-terphenyl ($C_{18}H_{14}$) which crystallises in the *P2₁/a* space group with herringbone packing (CCDC 1269381). Among organic crystals, p-terphenyl is relatively stable under electron beam exposure, with a critical fluence (CF) >330 e⁻ Å⁻² (Supplementary Figure 3). Figure 2a shows an annular dark field (ADF) STEM image of a p-terphenyl thin film with dense dislocation network. The length of individual dislocation lines that extend across the thin film, e.g. >2 μm for the selected dislocation (solid line), indicates that these dislocations are predominantly in-plane or <3° out-of-plane (estimated film thickness as 10-100 nm). The density of these dislocations in the field of view is in the range 0.32-3.2 x $10^{10}$ cm$^{-2}$ for the estimated thickness of the film (Supplementary Table 1). The area-averaged electron diffraction pattern was indexed to [001], a typical film orientation for p-terphenyl[20]. Figure 2c presents the recorded bend contour displacements as a function of azimuthal angle φ in a polar plot (φ is the angle relative to the horizontal axis of the diffraction patterns, see Supplementary Note 1). A fit to equation 1 is overlaid, showing correspondence between the experimental data and the functional form of the geometric model. Supplementary Note 2 and Supplementary Fig. 4-5 provide technical details of the curve fitting process, and Supplementary Table 2 includes fitting parameters. No displacement is visible for $\mathbf{g}_{200}$ or $\mathbf{g}_{\bar{2}00}$ while significant displacements are visible for other $\mathbf{g}_{hkl}$, e.g. $\mathbf{g}_{210}$ (Fig. 2c). Based on the extended, in-plane character of the dislocation we show that $\mathbf{B} = [010]$ (see also Methods) and identify the operative slip system as [010](001), consistent with previously inferred slip systems in p-terphenyl[20]. Given **B** is neither parallel nor perpendicular to **u**, the dislocation has mixed character. Supplementary Figure 6 shows two further dislocations from the same field of view, likewise assigned as mixed, $\mathbf{B} = [010]$ dislocations. The bend contours show the expected node pattern for screw components near $\mathbf{g}_{hkl} \cdot \mathbf{B}_{screw} = n$ (Supplementary Note 3)[27].

Critically, the fitting approach (equation 1) means SED-based dislocation analysis is not limited to areas or samples with a large number of $\mathbf{g}_{hkl}$. Practically, many samples may have a limited field of view, reducing the number of bend contours that can be used in analysis, or crystals may be tilted away from high symmetry zone axes, likewise reducing the number of $\mathbf{g}_{hkl}$ available for analysis. Supplementary Figures 7-9 present further analyses in p-terphenyl with incomplete sampling of equation 1 and for tilts estimated at 3.6° from [001] (Supplementary Figure 8), identifying additional Burgers vectors $\mathbf{B} = [110]$ and $[1\bar{2}0]$. These additional Burgers vectors are consistent with all slip systems reported previously by inference from TEM imaging[20]. We further demonstrate determination of the relative handedness of screw components between multiple dislocations by analysing the 'twisting' of the bend contours at the dislocations (Supplementary Figure 10).

Figure 3 highlights the resilience of the fitting approach, now applied to anthracene a much more beam sensitive material with CF ~20 e$^-$ Å$^{-2}$ (Supplementary Figure 11). Figure 3a depicts the anthracene unit cell (*P2$_1$/a*, CCDC 1103062), and Figure 3b presents an ADF-STEM image and corresponding dislocation network determined from the diffraction contrast. A large bend contour is visible at the top of Fig. 3b. No single VDF generated from this dataset showed a zero-displacement bend contour (as expected for $\mathbf{g}_{hkl} \cdot \mathbf{B} = 0$), as the magnification used to acquire this dataset meant many bend contours did not cross the dislocation lines within the field of view. Nevertheless, fitting of the polar plot is unambiguous (Fig. 3d) and supports determination of **B** = [010] for this mixed-type dislocation (Fig. 3e). Depending on the particular $\mathbf{g}_{hkl}$ bend contours crossing a dislocation line and the width of these contours, the approach results in a loss of precision in the Burgers vector direction. Supplementary Figure 12 illustrates such a case for a dislocation in anthracene where a low-index Burgers vector assignment is less definitive. In this case, the Burgers vector extracted from the polar function is a minimum 3.6° from a candidate **B** in the *ab* plane [*uv*0] = [$\bar{1}\bar{5}0$]. Burgers vectors in the *ab* plane reflect slip between the planes of anthracene molecules where intermolecular interactions are weakest. While this dislocation (Supplementary Figure 12) exhibits predominantly edge-type character, the relative handedness of the screw component is also discernible. In addition to dislocations in solution-grown anthracene films, analysis of dislocations in spin-coated films demonstrate the versatility of the method (Supplementary Figure 13). Photodimerisation and trap states have been linked to dislocations in anthracene[29,30], motivating further examination of individual dislocations in anthracene and wider organic optoelectronic materials.

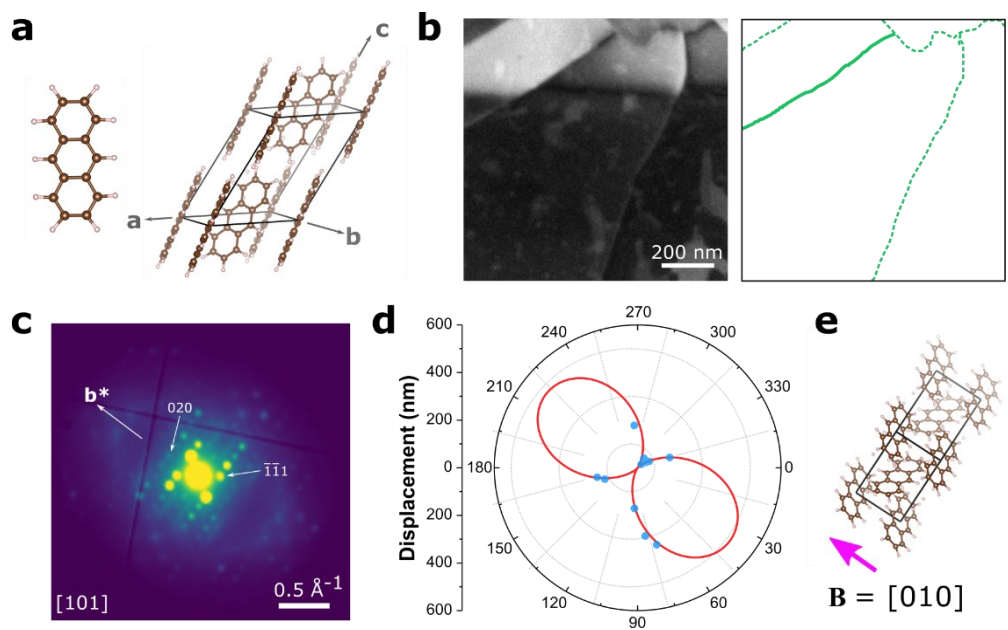

**Figure 3. Low-dose dislocation analysis in anthracene. a**, Representations of the anthracene molecule and unit cell. **b**, ADF-STEM image and corresponding extraction of the dislocation network from the diffraction contrast. **c**, Averaged electron diffraction pattern of the field of view in **b**, indexed to the [101] zone axis. **d**, A polar plot of the bend contour displacement as a function of azimuthal angle φ for the solid green dislocation line highlighted in **b**. The red trace marks a fit to equation 1 with the measured displacements in blue. **e**, The anthracene unit cell oriented to match the sample orientation. The magenta arrow marks the Burgers vector **B** = [010].

The versatility of our approach is further demonstrated in plate- crystals, rather than films, of theophylline, a model pharmaceutical compound (and a channel hydrate), and *n*-hentriacontane ($C_{31}H_{64}$), a paraffin wax component in leaf waxes and a model for wider long-chain hydrocarbon materials (Figure 4). Identification of crystallographic details of defects offers insights into their role in compaction and tableting[31] and in hydration/rehydration pathways on storage of drug powders[32] and in diffusion pathways for transpiration[33]. The CF for the stable theophylline polymorph (Form II) has been reported[6] within the range 26-36 e$^-$ Å$^{-2}$, and we have measured the CF for *n*-hentriacontane as 6 e$^-$ Å$^{-2}$, in line with previous reports of a CF between 5-10 e$^-$ Å$^{-2}$ for paraffin waxes[17]. Theophylline exhibits a number of polymorphs and hydrates, and we first show an example from theophylline anhydrous Form II (Fig. 4a-e) which crystallises in *Pna2$_1$*. Figure 4b shows a triangular plate crystal, with its electron diffraction pattern indexed to the [141] zone axis (Fig. 4c). In this crystal, the dislocation is short, which may indicate the dislocation line is not parallel to the plate surfaces. Still, VDF analysis of the available $g_{hkl}$ enables fitting to equation 1 (Fig. 4d). The obtained Burgers vector does not align with an in-plane low-index direction, but the vector does align with the projection of **B** = [102] which was therefore assigned as the Burgers vector with mixed character. Additional screw-type dislocations with **B** = [100] were identified in the monohydrate crystal structure (Form M, Supplementary Figure 14), as well as in the metastable theophylline anhydrous Form III (Supplementary Figure 15) which may play a role in the dehydration pathway of Form M to Form II[34]. These assignments indicate out-of-plane dislocations can be evaluated with the SED bend contour approach.

Figure 4e depicts *n*-hentriacontane and its corresponding *Pbcm* unit cell, modelled after the unit cells of other odd long-chain hydrocarbons[35]. As for theophylline, these crystals were plates, with a section depicted in Fig. 4f, oriented along the [001] zone axis (Fig. 4g). In this case, the constructed VDFs included $g_{hkl}$ exactly at the invisibility criterion as confirmed by fitting to equation 1 (Fig. 4h). Here, the dislocation is unambiguously identified with the Burgers vector direction along **B** = [010]. The dislocation line illustrates a full range of edge, screw, and mixed character along its curved path, with the length of the dislocation line suggesting the dislocation lies in (001), contrasting with screw dislocations with **B** along the long molecular axis reported previously in this class of materials[3]. Additional pure screw dislocations were recorded with the same Burgers vector direction (Supplementary Figure 16). While edge dislocations have been imaged in helium-cooled paraffins[17], basal plane mixed and screw dislocations, contrasting with spiral growth step dislocations with **B** along the long molecular axis in linear hydrocarbons[3], can now be readily identified in these highly beam-sensitive materials.

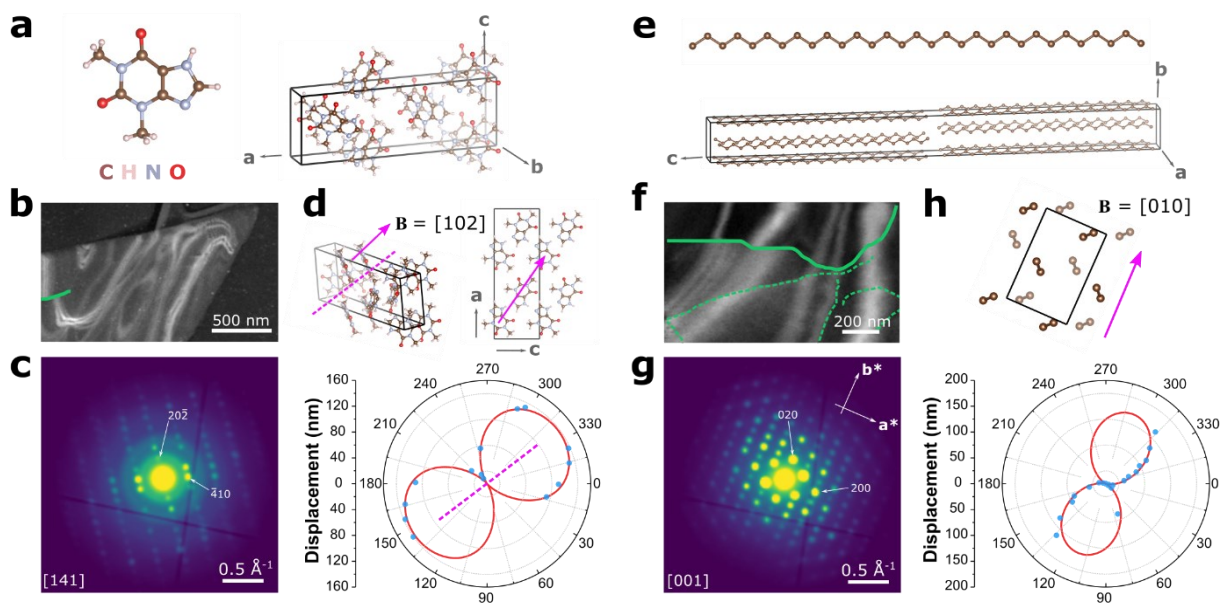

**Figure 4. Low-dose dislocation analysis in theophylline and paraffin crystals. a**, Representations of the theophylline molecule and unit cell of theophylline Form II. **b**, ADF-STEM image with a single dislocation line overlaid in green. **c**, The corresponding electron diffraction pattern, indexed to the [141] zone axis orientation. **d**, Burgers vector determination by fitting a polar plot of the bend contour displacement as a function of azimuthal angle φ to equation 1 for the dislocation line highlighted in **b**. The measured displacements are shown in blue with the fit shown in red. A dashed magenta line marks the experimentally determined Burgers vector direction **B** = [1 $\overline{16}$ 4] which aligns in projection to within ~3° with the assigned vector **B** = [102]. The unit cell is depicted at the same orientation as given by the diffraction pattern in **c** as well as along the [010] direction to better visualise the assigned Burgers vector. **e**, Representations of a *n*-hentriacontane ($C_{31}H_{64}$) molecule and unit cell. **f**, ADF-STEM image with dislocation lines marked in green. **g**, The corresponding electron diffraction pattern, indexed to the [001] zone axis orientation. **h**, Burgers vector determination by fitting a polar plot of the bend contour displacement as a function of azimuthal angle φ to equation 1 for the dislocation line highlighted in **f**. The measured displacements are shown in blue with the fit shown in red. The unit cell is depicted at the same orientation as given by the diffraction pattern in **g**. The magenta arrow follows the long axis of the lobes in the polar plot, giving the Burgers vector direction as **B** = [010].

In developing this approach, we note a few artefacts and limitations. During beam rastering, samples may abruptly change orientation which can produce linear features in diffraction contrast imaging, but these changes are readily distinguished from dislocation contrast (Supplementary Figure 17). Samples containing multiple finely spaced dislocation lines present challenges due to the finite diffraction-limited spatial resolution of ~ 3 nm when working with a low convergence semi-angle electron beam (<1 mrad). Further extensions to the suggested approach may be possible to account for the combined bend contour shift where a quasi-invisibility condition is still observed (Supplementary Figure 18-19), though we note this may require prior knowledge of contributing Burgers vectors, and we cannot rule out deviations from linear, elastic behaviour in the regions between such dislocation lines.

In summary, we have demonstrated dislocation analysis in varied organic molecular crystals from low-dose, single-exposure SED measurements. These dislocations exhibit characteristic Burgers vector directions that lie predominantly in planes containing weak intermolecular interactions between layers of molecules. These planes are likely to be those suited to insertion or deletion of nonparallel planes without cutting across the internal molecular structure. Identification of the specific, observed Burgers vector directions in these slip planes now provides input for computational modelling of experimentally constrained structures with known edge, screw, or mixed dislocation character. Moreover, the generalisation of this approach provides a means for routine analysis of extended dislocations in molecular crystals with CFs down to a few e$^-$ Å$^{-2}$, driving forward research into methods for inhibiting or designing dislocation formation during crystal growth and controlling their contributions to the functional properties of organic crystalline materials.

**Methods**

Sample preparation

Thin films of p-terphenyl and anthracene were prepared by solvent evaporation from xylene solutions[20]. Briefly, p-terphenyl (99.5%, Merck) or anthracene (99.0%, Merck) powder was dissolved in xylene to obtain a clear solution (~0.45-0.50 mg.mL$^{-1}$). 0.4 mL of the xylene solution was left to evaporate slowly on the surface of deionized water in a glass dish covered with Al foil to slow the rate of evaporation. Single-crystal films were obtained after two days of evaporation followed by half a day of aging. Interference colours were visible indicators of crystal thickness. Films that appeared blue or gold were sufficiently thin (120-200 nm) for electron microscopy[36], and these films were selected for the SED experiments. The single-crystal films were transferred onto lacey carbon film TEM support grids (EM Resolution) by placing the grid under the film and lifting to catch the film on the TEM grid. Prior to the transfer process, the TEM grids were treated with an Oxygen-Argon (25:75) plasma for 10 s using an HPT-100 plasma treatment system (Henniker Plasma). Grids were allowed to dry in ambient air. Spin-coated anthracene films were prepared on SiN$_x$ membrane window TEM grids (3x3 SiN$_x$ windows, 30 nm thickness, Norcada) exposed to an Ar plasma for a minute. For spin-coating, a 5-6 mg.mL$^{-1}$ solution of anthracene in chloroform was used. A SiN$_x$ membrane window was fixed on a custom chuck using tape and 30 μL of the casting solution was dropped on the membrane window at 200 rpm for 10 s. The spinning was then increased to 3000 rpm and the film was spun for a further minute. Finally, the SiN$_x$ grid was annealed at 80 °C on a hot plate for 15 minutes.

Single-crystal films of theophylline were prepared by the solvent evaporation approach[37,38]. Briefly, theophylline powder was dissolved in nitromethane followed by heating to ~60 °C to obtain a saturated solution. The solution was then drop-cast onto a holey carbon film TEM support grid (EM resolutions) and left to evaporate. Highly faulted particles of partially hydrated

Form II plates were produced by immersion in water. A sample of theophylline monohydrate (Form M) was prepared by evaporative crystallisation from either an ethanol:water or nitromethane:water solutions at 3:2 molar ratio followed by heating to ~55 °C. The solution was allowed to cool slowly to ~35 °C under continuous stirring for a period of 3 h. The solution was then drop-cast onto a holey carbon film TEM support grid and left to evaporate. Form III (indexed to Form IIIb structure[34]) was identified on a TEM grid with theophylline Form M (recrystallized from ethanol:water) and then immersed in liquid nitrogen. Data on form IIIb was likewise acquired under liquid nitrogen cooling (<110 K) using a Fischione cryo-transfer holder. All other reported data were acquired at ambient conditions. Single-crystal films of *n*-hentriacontane were likewise prepared by solvent evaporation[39] by drop-casting a supersaturated hexane solution (2 μL) onto a continuous carbon film TEM support grid (EM resolutions).

Transmission electron microscopy (TEM)

Electron diffraction pattern series as a function of cumulative electron fluence for p-terphenyl and anthracene were recorded using an FEI Titan[3] Themis 300 (X-FEG high-brightness electron source, operated at 300 kV) microscope in a standard parallel beam TEM configuration (Supplementary Note 5). For p-terphenyl an electron beam flux of 2.5 e$^-$ Å$^{-2}$ s$^{-1}$ was used whereas for anthracene an electron beam flux of 0.085 e$^-$ Å$^{-2}$ s$^{-1}$ was used. These values were determined from the current read out at the viewing screen, calibrated using a Faraday cup. A time-series of the selected area electron diffraction patterns were recorded for each compound across multiple crystals and multiple areas on the TEM grid.

Scanning electron diffraction (SED)

SED data were acquired on the JEOL ARM300CF instrument (ePSIC, Diamond Light Source, UK) equipped with a high-resolution pole piece, a cold field emission gun, aberration-correctors in both the probe-forming and image-forming optics, and a 4-chip Merlin/Medipix pixelated electron counting STEM detector. The instrument was operated at 300 kV. Selected samples were analysed at 200 kV, identified in the corresponding figure captions. The electron optics were configured for nanobeam diffraction by switching off the aberration corrector in the probe-forming optics and adjusting the condenser lens system to produce a convergence semi-angle of 0.8 mrad using a 10 μm condenser aperture. At 300 kV, this produces a 3-nm diffraction-limited probe diameter $d_{diff} = 1.22\lambda/\alpha$ for electron de Broglie wavelength *λ* and convergence semiangle *α*. The probe current was measured using a Faraday cup as ~1 pA, and the exposure time at each probe position was set as 1 ms. The electron fluence was approximately 8.8 e$^-$ Å$^{-2}$ at 300 kV or 5.4 e$^-$ Å$^{-2}$ at 200 kV in a single scan, assuming a disk-like probe with a diameter equal to $d_{diff}$. All SED measurements were conducted over a scan size of 256×256 probe positions. Image and diffraction calibration data, including reference calibration of residual elliptical distortion in the diffraction plane, were acquired using a gold diffraction cross-grating with a period of 500 nm (Ted Pella). The relative rotation between the diffraction pattern and the raster pattern was calibrated using standard $MoO_3$ crystals (Agar Scientific). Calibration data were acquired under identical conditions to the molecular crystal samples.

Data analysis

SED data were processed, aligned, and calibrated using Pyxem-0.11.0 and supporting tools from the HyperSpy package (1.6.5)[40] according to previously reported procedures[41]. The direct beam in each diffraction pattern was aligned and moved to the pattern centre using a cross-correlation function in Pyxem followed by application of an affine transformation matrix to correct the elliptical distortion and elimination of the rotation offset between the SED raster pattern and the diffraction data. Annular Dark Field (ADF) images were formed by integrating the diffraction pattern at each probe position between an inner radius 0.12 Å$^{-1}$ ($2\theta$ = 2.7 mrad) and an outer radius 1 Å$^{-1}$ ($2\theta$ = 22.4 mrad) to produce an image dominated by diffraction contrast. Average electron diffraction patterns were obtained by taking the mean intensity from all diffraction patterns contributing to an image. Diffraction patterns are presented as the square-root of recorded intensity (applied in ImageJ software) for improved visualisation of low- and high-intensity features.

Diffraction patterns were indexed using CrystalMaker software. Simulated kinematical electron diffraction patterns were generated from the CIF files for p-terphenyl (CCDC 1269381), anthracene (CCDC 1103062), theophylline (CCDC 128707 for Form II, CCDC 183780 for Form M, and Ref. [34] for metastable form IIIb). A unit cell for *n*-hentriacontane ($C_{31}H_{64}$) was constructed from previously reported odd number, long-chain hydrocarbons[35]. Diffraction patterns at different major orientations for each material were compared to the calibrated experimental diffraction patterns in overlay. In most cases, the samples were indexed to low-index zone axes such as [001] for p-terphenyl, [101] for anthracene, and [001] for *n*-hentriacontane. Mis-tilt from high symmetry zone axes, e.g. Supplementary Figure 8, was estimated by inspection of simulated diffraction patterns tilted away from the high symmetry zone axis to match experimental data. For theophylline, auto-indexing in CrystalMaker was used to assist in finding the possible zone axes as these samples could not be indexed to low-index zone axes. All simulated diffraction patterns at candidate zone axes were inspected and compared to the calibrated experimental diffraction patterns. Following this approach, we indexed the theophylline crystals to the zone axes [141] for Form II theophylline, [$\bar{2}$13] for Form M theophylline, and [211] for Metastable Form III.

Automated constructions of multiple virtual dark field (VDF) images was scripted with Python using the functions from Pyxem-0.11.0 and Hyperspy-1.6.5 packages. Diffraction patterns, originally 515×515 pixels, were cropped and then binned by a factor of two to produce diffraction patterns with dimensions 257×257 pixels. Bragg disks were detected using a peak finding function using the difference of Gaussians method. Filter settings were tuned iteratively and assessed by manual inspection to capture all disk-like diffraction features in a randomized sub-sample of diffraction patterns. After peak finding, the position of each Bragg disk in the diffraction pattern was recorded in diffraction vector ($k_x$, $k_y$) coordinates. The peak position was further refined using a local centre of mass search. VDF images were then constructed from the integrated intensity within a circular aperture centred at the peak position for each probe position in the SED scan. The aperture radius was adjusted to minimise Bragg disk overlap, typically set in the range 0.016-0.026 Å$^{-1}$. The angle of the associated diffraction vector **g**$_{hkl}$ for each Bragg disk taken as the inverse tangent of $k_y/k_x$.

Bend contour displacements were measured in ImageJ. The displacement of the bend contours on crossing the dislocation line is determined as the distance, tracing along the dislocation line, between the middle of the two bend contours, as highlighting by the red dashed lines in Supplementary Figure 1 and Supplementary Note 1. The azimuthal angular variation in bend contour displacements was analysed by nonlinear least squares curve fitting, following equation 1. Curve fitting was implemented using the curve_fit function in the SciPy Python package with further evaluation of confidence intervals using the Lmfit-1.2.1 Python package[42] (Supplementary Note 4).

## Acknowledgements

STP and SMC acknowledge funding support from the UK's Engineering and Physical Sciences Research Council (EPSRC, EP/V044907/1). EW, AB, and SMC acknowledge funding for a studentship supported by the EPSRC Centre for Doctoral Training in Molecules to Product (EP/SO22473/1), in collaboration with Syngenta and AstraZeneca. NK and AB acknowledge funding from AstraZeneca Ltd and the EPSRC for an iCASE studentship (No 2182593). We thank the Diamond Light Source, Rutherford Appleton Laboratory, UK, for access to the electron Physical Sciences Imaging Centre (ePSIC, MG28500, MG30057, MG31258, MG30157, and MG31872). We thank Mohsen Danaie, David Hopkinson, and Christopher Allen for support at ePSIC.

# References

1. Burton, W. K., Cabrera, N., Frank, F. C. & Mott, N. F. The growth of crystals and the equilibrium structure of their surfaces. *Philosophical Transactions of the Royal Society of London. Series A, Mathematical and Physical Sciences* **243**, 299–358 (1997).
2. Zhong, X., Shtukenberg, A. G., Hueckel, T., Kahr, B. & Ward, M. D. Screw Dislocation Generation by Inclusions in Molecular Crystals. *Crystal Growth & Design* **18**, 318–323 (2018).
3. Olson, I. A., Shtukenberg, A. G., Kahr, B. & Ward, M. D. Dislocations in molecular crystals. *Rep Prog Phys* **81**, 096501 (2018).
4. Jing, Y., Zhang, Y., Blendell, J., Koslowski, M. & Carvajal, M. T. Nanoindentation Method To Study Slip Planes in Molecular Crystals in a Systematic Manner. *Crystal Growth & Design* **11**, 5260–5267 (2011).
5. Haneef, H. F., Zeidell, A. M. & Jurchescu, O. D. Charge carrier traps in organic semiconductors: a review on the underlying physics and impact on electronic devices. *J. Mater. Chem. C* **8**, 759–787 (2020).
6. Ilett, M. *et al.* Analysis of complex, beam-sensitive materials by transmission electron microscopy and associated techniques. *Phil. Trans. Roy. Soc. A* **378**, 20190601 (2020).
7. Hirsch, P. B., Horne, R. W. & Whelan, M. J. LXVIII. Direct observations of the arrangement and motion of dislocations in aluminium. *The Philosophical Magazine: A Journal of Theoretical Experimental and Applied Physics* **1**, 677–684 (1956).
8. Lee, C. *et al.* Strength can be controlled by edge dislocations in refractory high-entropy alloys. *Nat Commun* **12**, 5474 (2021).
9. Schubert, M. F. *et al.* Effect of dislocation density on efficiency droop in GaInN∕GaN light-emitting diodes. *Appl. Phys. Lett.* **91**, 231114 (2007).
10. Armstrong, M. D., Lan, K.-W., Guo, Y. & Perry, N. H. Dislocation-Mediated Conductivity in Oxides: Progress, Challenges, and Opportunities. *ACS Nano* **15**, 9211–9221 (2021).
11. Plomp, M., van Enckevort, W. J. P., Hoof, P. J. C. M. van & Streek, C. J. van de. Morphology of and dislocation movement in n-C40H82 paraffin crystals grown from solution. *Journal of Crystal Growth* **249**, 600–613 (2003).
12. Klapper, H. X-Ray Topography of Organic Crystals. in *Organic Crystals I: Characterization* (ed. Karl, N.) 109–162 (Springer, 1991). doi:10.1007/978-3-642-76253-6_3.
13. Chen, C.-C. *et al.* Three-dimensional imaging of dislocations in a nanoparticle at atomic resolution. *Nature* **496**, 74–77 (2013).
14. Rothmann, M. U. *et al.* Atomic-scale microstructure of metal halide perovskite. *Science* **370**, (2020).
15. Barnard, J. S., Sharp, J., Tong, J. R. & Midgley, P. A. High-Resolution Three-Dimensional Imaging of Dislocations. *Science* **313**, 319–319 (2006).
16. Yang, H. *et al.* Imaging screw dislocations at atomic resolution by aberration-corrected electron optical sectioning. *Nat Commun* **6**, 7266 (2015).
17. Zemlin, F., Reuber, E., Beckmann, E., Zeitler, E. & Dorset, D. L. Molecular Resolution Electron Micrographs of Monolamellar Paraffin Crystals. *Science* **229**, 461–462 (1985).
18. Drummy, L. f., Kübel, C., Lee, D., White, A. & Martin, D. c. Direct Imaging of Defect Structures in Pentacene Nanocrystals. *Advanced Materials* **14**, 54–57 (2002).
19. Jones, W. & Williams, J. O. Real space crystallography and defects in molecular crystals. *J Mater Sci* **10**, 379–386 (1975).
20. Jones, W., Thomas, J. M., Williams, J. O. & Hobbs, L. W. Electron microscopic studies of extended defects in organic molecular crystals. Part 1.—p-Terphenyl. *J. Chem. Soc., Faraday Trans. 2* **71**, 138–145 (1975).
21. Parkinson, G. M. & Davies, E. Electron Microscopic Identififcation and Chemical Consequences of Extended Defects in Organic Molecular Crystals. *J. Phys. Colloques* **39**, C2-67 (1978).
22. Bustillo, K. C. *et al.* 4D-STEM of Beam-Sensitive Materials. *Acc. Chem. Res.* (2021) doi:10.1021/acs.accounts.1c00073.

23. Gallagher-Jones, M. *et al.* Nanoscale mosaicity revealed in peptide microcrystals by scanning electron nanodiffraction. *Communications Biology* **2**, 1–8 (2019).
24. Sneyd, A. J. *et al.* Efficient energy transport in an organic semiconductor mediated by transient exciton delocalization. *Science Advances* **7**, eabh4232 (2021).
25. Panova, O. *et al.* Diffraction imaging of nanocrystalline structures in organic semiconductor molecular thin films. *Nat. Mater.* **18**, 860–865 (2019).
26. Balhorn, L. *et al.* Closing the loop between microstructure and charge transport in conjugated polymers by combining microscopy and simulation. *Proceedings of the National Academy of Sciences* **119**, e2204346119 (2022).
27. Spiecker, E. & Jäger, W. Burgers vector analysis of large area misfit dislocation arrays from bend contour contrast in transmission electron microscope images. *J. Phys.: Condens. Matter* **14**, 12767–12776 (2002).
28. Cherns, D. & Preston, A. R. Convergent beam diffraction studies of interfaces, defects, and multilayers. *Journal of Electron Microscopy Technique* **13**, 111–122 (1989).
29. Williams, J. O. & Zboiński, Z. Structural imperfections and the delayed fluorescence of anthracene crystals. *J. Chem. Soc., Faraday Trans. 2* **74**, 618–629 (1978).
30. Williams, J. O. & Thomas, J. M. Photochemical Reactions Inside the Electron Microscope: Preferred Dimerization of Anthracene at Dislocations. *Molecular Crystals and Liquid Crystals* **16**, 371–375 (1972).
31. Jain, S. Mechanical properties of powders for compaction and tableting: an overview. *Pharmaceutical Science & Technology Today* **2**, 20–31 (1999).
32. Amado, A. M., Nolasco, M. M. & Ribeiro-Claro, P. J. A. Probing Pseudopolymorphic Transitions in Pharmaceutical Solids using Raman Spectroscopy: Hydration and Dehydration of Theophylline. *JPharmSci* **96**, 1366–1379 (2007).
33. *Biology of the Plant Cuticle (Annual Plant Reviews): 23*. (Wiley-Blackwell, 2006).
34. Paiva, E. M. *et al.* Understanding the Metastability of Theophylline FIII by Means of Low-Frequency Vibrational Spectroscopy. *Mol. Pharmaceutics* **18**, 3578–3587 (2021).
35. Smith, A. E. The Crystal Structure of the Normal Paraffin Hydrocarbons. *J. Chem. Phys.* **21**, 2229–2231 (1953).
36. Howie, A., Rocca, F. J. & Valdrè, U. Electron beam ionization damage processes in p-terphenyl. *Philosophical Magazine B* **52**, 751–757 (1985).
37. Eddleston, M. D. *et al.* Highly Unusual Triangular Crystals of Theophylline: The Influence of Solvent on the Growth Rates of Polar Crystal Faces. *Crystal Growth & Design* **15**, 2514–2523 (2015).
38. Zhu, H., Yuen, C. & Grant, D. J. W. Influence of water activity in organic solvent + water mixtures on the nature of the crystallizing drug phase. 1. Theophylline. *International Journal of Pharmaceutics* **135**, 151–160 (1996).
39. Fryer, J. R., McConnell, C. H., Dorset, D. L., Zemlin, F. & Zeitler, E. High resolution electron microscopy of molecular crystals. IV. Paraffins and their solid solutions. *Proceedings of the Royal Society of London. Series A: Mathematical, Physical and Engineering Sciences* **453**, 1929–1946 (1997).
40. Peña, F. de la *et al.* hyperspy/hyperspy: Release v1.6.5. (2021) doi:10.5281/zenodo.5608741.
41. Duncan N. Johnstone *et al.* pyxem/pyxem-demos: pyxem-demos 0.11.0. (2020) doi:10.5281/zenodo.3831456.
42. Newville, M. *et al.* lmfit/lmfit-py: 1.2.1. (2023) doi:10.5281/zenodo.7887568.

# Supplementary Information

**Microscopic crystallographic analysis of dislocations in molecular crystals**

Sang T. Pham[1], Natalia Koniuch[1], Emily Wynne[1],
Andy Brown[1], Sean M. Collins[1,2]*

[1]*Bragg Centre for Materials Research & School of Chemical and Process Engineering, University of Leeds, Woodhouse Lane, Leeds LS2 9JT, UK*
[2]*School of Chemistry, University of Leeds, Woodhouse Lane, Leeds LS2 9JT, UK*

*Email: s.m.collins@leeds.ac.uk

## Contents

**Supplementary Note 1: Bend contour method for Burgers vector analysis**

*Measurements of bend contour displacement and azimuthal angle*

Bend contour displacements were measured as the distance along the dislocation core between two lines marking the central path of the bend contour, with bend contours typically appearing as approximately parallel lines on either side of the dislocation core. In cases where twisting of the bend contour at the dislocation line was present, the middle of the bend contour was determined away from dislocation core (two red dashed lines in Supplementary Figure 1b) to avoid overestimation of the displacement measurement by the short-range deflections of the bend contour ('twisting') caused by the dislocation strain field. Supplementary Figure 1c shows a challenging case where no gap in intensity appears, but an offset at the dislocation core was still observable. Moreover, this example was taken from the edge of a field of view with only part of the bend contour to the right of the dislocation core included in the dataset. As such, the bend contour to the right of the dislocation line was taken as parallel to the bend contour to the left (as for Supplementary Figure 1b), but with the centre estimated at the right-hand edge of the field of view. Such cases may introduce some loss of precision in the displacement measurements but did not preclude construction of a polar plot for analysis of the dislocation Burgers vector (see also Supplementary Figure 12).

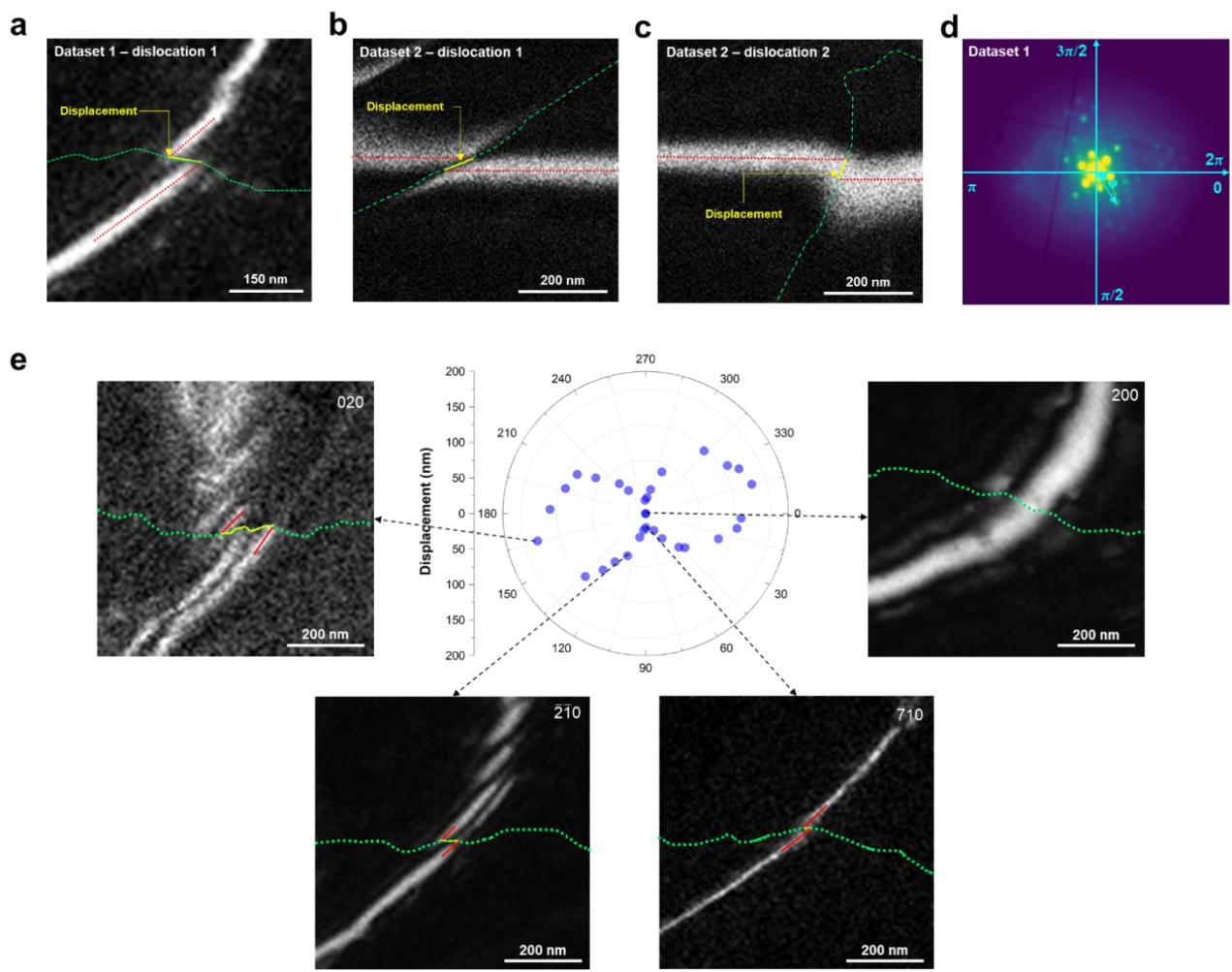

**Supplementary Figure 1.** Measurements of bend contour displacements. **a**, An example in a p-terphenyl film showing a symmetrical break of the bend contour along the dislocation line (Dataset 1 is also shown in Fig. 2). **b**, An example in an anthracene dataset showing a symmetrical break and twisting of the bend contour along the dislocation line (Dataset 2 is also shown in Fig. 3, denoted as dislocation-1) **c**, An example in an anthracene dataset showing displacement of the bend contour direction on passing the dislocation line without a gap in the intensity profile (Dataset 2 as in **b**; this dislocation is denoted dislocation-2 and is also analysed in Supplementary Figure 12). **d**, Illustration of the definition of the azimuthal angle φ as the angle between a selected **g**$_{hkl}$ vector and the horizontal axis of the diffraction pattern (shown using the Dataset 1 average pattern, also presented in Fig. 2). **e**, VDF images formed by placing virtual apertures at different **g**$_{hkl}$ showing an increasing trend in the bend contour displacement when moving away from **g**$_{200}$.

*Geometric model of bend contour displacement*

We introduce a geometric solution for the bend contour displacement on crossing dislocations in Supplementary Figure 2. Simple representation of the atomic displacement via **B** direction is shown for edge-type dislocation (Supplementary Figure 2a) and screw-type dislocation (Supplementary Figure 2b). From the viewing projection along the Cartesian *z*-axis for both dislocation types defined in the *xy*-plane, a trigonometric relationship can be derived between the Burgers vector **B** and the angle between **B** and the plane normal (**g**$_{hkl}$):

$$x = \chi \mathbf{B} \cos^2(\varphi) \tag{S-1}$$

with $\chi\mathbf{B}$ is a constant for any fixed set of displacements for a given dislocation. For a perfect crystal with no defects presented, local bending of the sample, at the angles larger than the Bragg angle, defines a set of diffraction conditions for a particular set of planes where the planes are brought into or through the Bragg condition. This condition produces bend contours running across images of the film. Putting it another way, the bend contour is formed by the tilting of the lattice planes, due to local bending, to an angle *θ*$_{hkl}$ along the **g**$_{hkl}$ to intersect with the Ewald sphere at zero excitation error. The strain field around a dislocation core produces additional local tilt of surrounding planes as 'seen' by the **g**$_{hkl}$. The additional tilts of planes at the core cause a shift in the bend contour at the dislocation line to match the new angle required to satisfy the Bragg condition exactly. Following the decomposition of the **g**$_{hkl}$ into the vector parallel to **B** (**g**$^{\|\mathbf{B}}$) and the vector perpendicular to **B** (**g**$^{\perp\mathbf{B}}$). It can be noticed that the component **g**$^{\|\mathbf{B}}$ changes under tilting due to the dislocation and the component **g**$^{\perp\mathbf{B}}$ is unchanged. To have the same angle as where **g**$_{hkl}$ intersects the Ewald sphere requires a change in **g**$^{\|\mathbf{B}}$ only which is then linked to the angle *θ*$^{\|\mathbf{B}}$, and is thus, linked to the parameter x. We can now calculate the *θ*$^{\|\mathbf{B}}$ for the specific *hkl* planes (Supplementary Figure 2c) as:

$$\theta^{\|\mathbf{B}}{}_{hkl} = \arctan\left(\frac{x}{d_\perp}\right) \tag{S-2}$$

with $d_\perp$ is the d-spacing that allows for completing the triangle to translate the x parameter into an angle *θ*$^{\|\mathbf{B}}$. The vertical R line is taken in Supplementary Figure 2d as a reference angle for the exact Bragg condition (bend contour criterion) while the red line and blue line indicate the tilted planes at either sides of the dislocation core. The shift from the vertical R line will be approximately the arc length R*θ* for tilt angle *θ* for local tilting caused by dislocation is *θ*$^{\|\mathbf{B}}$. Hence, the total shift between the red and blue tilted planes at the dislocation core will be ~2R*θ*$^{\|\mathbf{B}}$. We can now construct the function for total shift/displacement of the bend contour on crossing the dislocation as:

$$f(\varphi) = 2R\theta^{\|\mathbf{B}} = 2R \arctan\left(\frac{x}{d_\perp}\right) = 2R \arctan\left(\frac{\chi\mathbf{B}\cos^2(\varphi)}{d_\perp}\right) \tag{S-3}$$

The function f(φ) can be simplified to: $f(\varphi) = A \arctan(B \cos^2(\varphi - C))$ where A, B, and C are fitting coefficients. From equation S-3, it can be noticed that f(φ) will be 0 when φ is 90°, i.e. the **g**$_{hkl}$ perpendicular to the Burgers vector ($\mathbf{g}_{hkl} \cdot \mathbf{B} = 0$), which is equivalent to when the Burgers vector lies parallel to the planes. It is in agreement with the definition for the invisibility criterion that states: the strain field around the dislocations causes local deviations or distortion of diffracting planes from their regular arrangement unless the Burgers vector of the dislocations lies parallel to the diffracting planes[1]. As such, the bend contours, associated to these planes, show no break on crossing the dislocation.

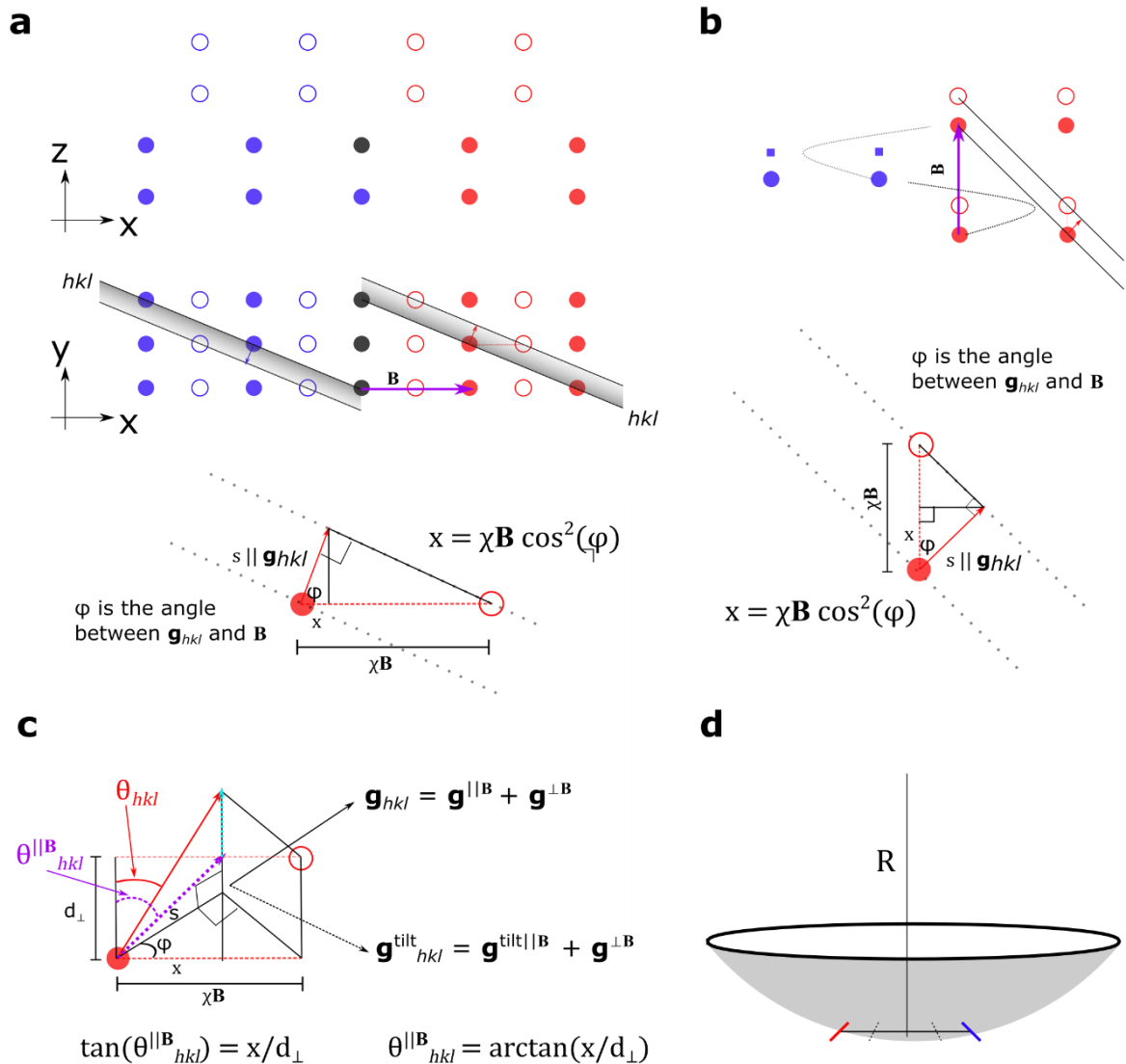

**Supplementary Figure 2. Geometry solution for lattice plane displacements due to dislocations in the bending crystals.** Geometric relationship between the Burgers vector and the angle φ via the component x with φ defined as the angle between **B** and plane normal $g_{hkl}$ for: **a**, edge-type dislocation and **b**, screw-type dislocation. The component x can be understood as the fraction of the displacement and associated local tilt at the dislocation core as 'seen' by $g_{hkl}$. **c**, Worked solution for the shift of a bend contour due to local tilts at a dislocation core locally changing the Bragg condition for $g_{hkl}$. **d**, Illustration of the displacement for a film section with radius of curvature R. The vertical R is taken as a reference angle for the exact Bragg condition (bend contour criterion). A large value of R corresponds to a flat (low curvature) surface; large displacements required to achieve angle adjustment. Displacements will generally be smaller than those shown in the schematic.

*Burgers vector determination*

The Burgers vector $\mathbf{B}$ is determined from the invisibility criterion condition $\mathbf{g}_{hkl} \cdot \mathbf{B} = 0$ where $\mathbf{g}_{hkl}$ at this condition can be found directly from the analysis of VDF images or can be estimated by the curve fitting using function $f(\varphi)$. For extended, in-plane dislocations ($\mathbf{g}_{hkl} \cdot (\mathbf{B} \times \mathbf{u}) = 0$), the Burgers vector direction can be recovered unambiguously. In this case, the Burgers vectors are also in-plane and perpendicular to the beam direction $\mathbf{z}$, i.e. $\mathbf{B} \times \mathbf{u} \parallel \mathbf{z}$. Thus, we can equivalently define the condition $\mathbf{B} \cdot \mathbf{z} = 0$ with $\mathbf{z}$ taken as the viewing direction (zone axis direction). For $\mathbf{B} = [u_1\ v_1\ w_1]$, $\mathbf{z} = [u_2\ v_2\ w_2]$, and $\mathbf{g} = (hkl)$, we obtain set of equations:

$$\begin{cases} hu_1 + kv_1 + lw_1 = 0 \\ (u_1\mathbf{a}+v_1\mathbf{b}+w_1\mathbf{c}) \cdot (u_2\mathbf{a}+v_2\mathbf{b}+w_2\mathbf{c}) = 0 \end{cases} \tag{S-4}$$

with $\mathbf{a}$, $\mathbf{b}$, and $\mathbf{c}$ are the unit cell translations in real space. Since $\mathbf{z}$ and $\mathbf{B}$ are represented in the lattice vector basis, $u_1$, $v_1$, $w_1$, $u_2$, $v_2$, and $w_2$ are all integers. Expanding the second equation in S-4, we have:

$$u_1 \cdot u_2 \cdot |\mathbf{a}|^2 + (u_1 v_2 + v_1 u_2) \cdot |\mathbf{a}| \cdot |\mathbf{b}| \cdot \cos(\gamma) + (u_1 w_2 + w_1 u_2) \cdot |\mathbf{a}| \cdot |\mathbf{c}| \cdot \cos(\beta)$$
$$+ v_1 \cdot v_2 \cdot |\mathbf{b}|^2 + (v_1 w_2 + w_1 v_2) \cdot |\mathbf{b}| \cdot |\mathbf{c}| \cdot \cos(\alpha) + w_1 \cdot w_2 \cdot |\mathbf{c}|^2 = 0 \tag{S-5}$$

We can now notice that equation S-5 can be simplified to $u_1 u_2 + v_1 v_2 + w_1 w_2 = 0$ if $\alpha = \beta = \gamma = 90°$ and $|\mathbf{a}| = |\mathbf{b}| = |\mathbf{c}|$, i.e. a cubic crystal. However, the studied molecular crystals in this study have non-cubic structure. As a result, equation S-5 must be used when calculating the Burgers vector $\mathbf{B}$. For given crystal structures, unit cell parameters, and zone axes from the studied materials, sets of equations to calculate $\mathbf{B}$ can be deduced as laid out below:

- For p-terphenyl (monoclinic) and *n*-hentriacontane (orthorhombic) viewed along $\mathbf{z} = [001]$, S-5 becomes:

$$\begin{cases} hu_1 + kv_1 + lw_1 = 0 \\ u_1 \cdot |\mathbf{a}| \cdot \cos(\beta) + w_1 \cdot |\mathbf{c}| = 0 \end{cases} \tag{S-6}$$

- For anthracene (monoclinic) viewed along $\mathbf{z} = [101]$, S-5 becomes:

$$\begin{cases} hu_1 + kv_1 + lw_1 = 0 \\ u_1 \cdot |\mathbf{a}|^2 + (u_1 + w_1) \cdot |\mathbf{a}| \cdot |\mathbf{c}| \cdot \cos(\beta) + w_1 \cdot |\mathbf{c}|^2 = 0 \end{cases} \tag{S-7}$$

- For theophylline form II (orthorhombic) viewed along $\mathbf{z} = [141]$, S-5 becomes:

$$\begin{cases} hu_1 + kv_1 + lw_1 = 0 \\ u_1 \cdot |\mathbf{a}|^2 + 4 \cdot v_1 \cdot |\mathbf{b}|^2 + w_1 \cdot |\mathbf{c}|^2 = 0 \end{cases} \tag{S-8}$$

- For theophylline form III (monoclinic) viewed along $\mathbf{z} = [211]$, S-5 becomes:

$$\begin{cases} hu_1 + kv_1 + lw_1 = 0 \\ 2u_1 \cdot |\mathbf{a}|^2 + (u_1 + 2w_1) \cdot |\mathbf{a}| \cdot |\mathbf{c}| \cdot \cos(\beta) + v_1 \cdot |\mathbf{b}|^2 + w_1 \cdot |\mathbf{c}|^2 = 0 \end{cases} \tag{S-9}$$

- For theophylline form M (monoclinic) viewed along $z = [\bar{2}13]$, S-5 becomes:

$$\begin{cases} hu_1 + kv_1 + lw_1 = 0 \\ -2u_1.|a|^2 + (3u_1 - 2w_1).|a|.|c|.\cos(\beta) + v_1.|b|^2 + 3w_1.|c|^2 = 0 \end{cases} \quad \text{(S-10)}$$

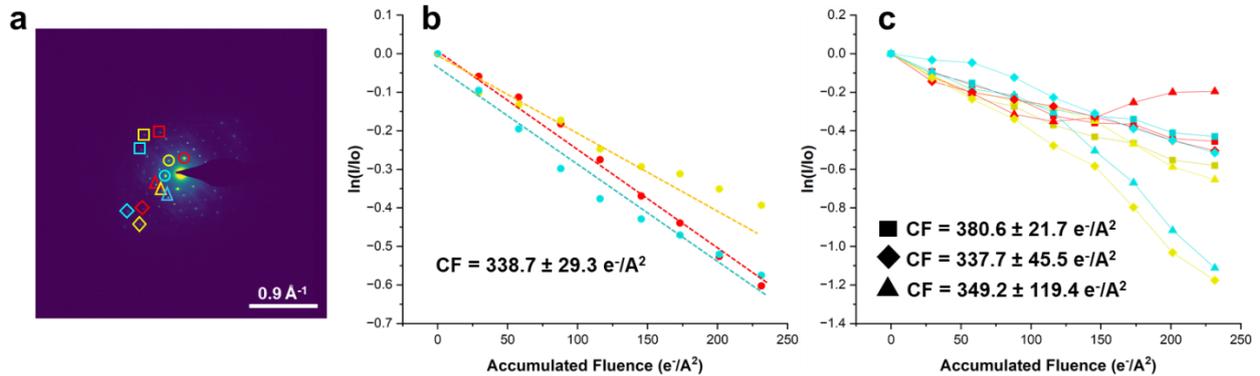

**Supplementary Figure 3. Loss of Bragg diffraction intensity with cumulative electron fluence (beam-damage series) for a p-terphenyl thin film viewed along [001]. a,** Selected Area Electron Diffraction Pattern (SAED) taken at the beginning of the experiment ($t_o$) showing the marked diffraction spots for CF analysis. **b**, Decay profile of the lattice planes having $d$-spacing > 4 Å. **c**, Decay profile of the lattice planes having $d$-spacing < 4 Å. The square-marked denote the diffraction spots corresponding to lattice planes with $d$-spacings 1.3-1.6 Å. The diamond-marked spots denote the diffraction spots corresponding to lattice planes with $d$-spacings 1.1-1.4 Å. The triangle-marked spots denote the diffraction spots corresponding to the lattice planes with $d$-spacings 2.6-2.9 Å. The decay profiles show an approximately log-linear response with adjusted determination coefficient $R^2 > 0.9$. The CF is extracted from the gradient of the linear fit (see also Supplementary Note 4).

**Supplementary Table 1. Estimated in-plane dislocation densities for molecular crystal samples.** The dislocation density is determined from measurement of the total length of all observed in-plane dislocations per unit volume of the crystal. The volume was estimated from the area of the field of view and an estimated sample thickness $T$, taken as a range from 10-100 nm. This estimate was consistent with thickness contrast from amorphous carbon support films (lacey carbon). Thickness likely varies within and between samples, and we report the dislocation densities here as order of magnitude estimates.

| Figure | Length (nm) | Area (nm$^2$) | $T$ (nm) | Volume (nm$^3$) | Dis. Density (cm$^{-2}$) (x10$^{10}$) | Sample |
|---|---|---|---|---|---|---|
| Fig. 2b | 12777 | 4.0 x 10$^6$ | 10-100 | 4.0x10$^7$-4.0x10$^8$ | 0.3-3.2 | p-terphenyl |
| Fig. 3b | 2594 | 10$^6$ | 10-100 | 10$^7$-10$^8$ | 0.3-2.6 | anthracene |
| Fig. 4b | 1040 | 2.0 x 10$^6$ | 10-100 | 2.0 x10$^7$-2.0x10$^8$ | 0.1-0.5 | theophylline |
| Fig. 4f | 3924 | 10$^6$ | 10-100 | 10$^7$-10$^8$ | 0.4-3.8 | *n*-hentriacontane |
| Supp. Fig. 7a | 1040 | 10$^6$ | 10-100 | 10$^7$-10$^8$ | 0.1-1.0 | p-terphenyl |
| Supp. Fig. 8a | 1640 | 10$^6$ | 10-100 | 10$^7$-10$^8$ | 0.2-1.7 | p-terphenyl |
| Supp. Fig. 9 | 10821 | 7.0 x 10$^6$ | 10-100 | 7.2x10$^7$-7.2x10$^8$ | 0.2-1.5 | p-terphenyl |
| Supp. Fig. 13a | 4221 | 4.0 x 10$^6$ | 10-100 | 4.0x10$^7$-4.0x10$^8$ | 0.1-1.1 | anthracene |
| Supp. Fig. 14a | 10560 | 1.5 x 10$^6$ | 10-100 | 1.5x10$^7$-1.5x10$^8$ | 0.7-6.9 | theophylline |
| Supp. Fig. 15a | 978 | 1.3 x 10$^6$ | 10-100 | 1.3x10$^7$-1.3x10$^8$ | 0.1-0.7 | theophylline |
| Supp. Fig. 16a | 13889 | 2.2 x 10$^6$ | 10-100 | 2.2x10$^7$-2.2x10$^8$ | 0.6-6.3 | *n*-hentriacontane |
| Supp. Fig. 18a | 1944 | 3.8 x 10$^6$ | 10-100 | 3.8x10$^7$-3.8x10$^8$ | 0.1-0.5 | p-terphenyl |
| Supp. Fig. 19a | 2040 | 0.5 x 10$^6$ | 10-100 | 0.5x10$^7$-0.5x10$^8$ | 0.4-4.5 | anthracene |

**Supplementary Note 2: Goodness-of-fit evaluation for polar function fitting**

We report the fitting function (equation 1) $f(\varphi) = A \arctan(B \cos^2(\varphi - C))$ arising from the geometric model for the bend contour displacement on crossing the dislocations (Supplementary Figure 2) to estimate the Burgers vector direction from experimental displacement measurements. We have evaluated the goodness-of-fit of the proposed model using goodness-of-fit statistics and together with inspection of the residuals, and evaluation of uncertainties in fitting parameter estimation. For goodness-of-fit statistics, the coefficient of determination $R^2$ and the chi-square ($\chi^2$) test statistic are used. Meanwhile, visualization of the fitting on the measured data and residual analysis are used as graphical methods to evaluate the fit.

*$R^2$ statistic*

The $R^2$ statistic seeks to summarise how close the fitted curve passes through the recorded measurements, most commonly used in linear regression analysis. $R^2$ can be defined from the sum of squared residuals ($SS_{res}$) and the total sum of squares ($SS_{tot}$):

$$R^2 = 1 - \frac{SS_{res}}{SS_{tot}} = 1 - \frac{\sum_i (y_i - f_i)^2}{\sum_i (y_i - \bar{y})^2} \tag{S-11}$$

with $y_i$ is the measured data, $f_i$ is the predicted data from the fitting model, and $\bar{y}$ is the mean of the measured data. Where the predicted values determined from the fitting model exactly match the measured values, the $SS_{res} = 0$ and $R^2 = 1$. We report this statistic as an indicator of the match between experiment and the fitted curve, while noting its limitations in distinguishing distributed errors arising from experimental uncertainty and systematic errors in a poor model correspondence to the recorded measurements.

*$\chi^2$ statistic*

The $\chi^2$ statistic offers a route to evaluate a fitted curve in the context of variance in experimental measurements. The simplest $\chi^2$ can be calculated as follows:

$$\chi^2 = \sum_i \frac{(y_i - f(x_i))^2}{\sigma_i^2} \tag{S-12}$$

where $y_i$ is again the measured data, $f(x)$ is the fitting model, $x_i$ is the independent variable, and $\sigma_i$ is the uncertainties/standard variation. For the purposes of the evaluation of curve fitting, the chi-square per degrees of freedom or reduced chi-square is useful as it takes into consideration the number of data points (measurements) and number of parameters describing the fitted function. The reduced chi-square can be calculated by dividing the chi-square to the number of degrees of freedom $\nu$:

$$\chi^2_{red} = \chi^2 / \nu \tag{S-13}$$

where $\nu = N - n_p$ for $N$ data points and $n_p$ parameters in the fitting model. A model (curve) that fits the data within the variance inherent in the data will exhibit a $\chi^2_{red}$ near 1. A $\chi^2_{red}$ much greater than 1 will indicate a poor agreement between the model and the measurements, and a $\chi^2_{red}$ less than 1 will indicate the model fits more closely to the data than expected for the inherent variation in the measurements.

To be able to calculate the $\chi^2$, and thus $\chi^2_{red}$, it is essential to estimate the variance $\sigma_i^2$ in the measured data, i.e. the variance in the measured bend contour displacements. The variance of the bend contour displacements likely differ for different bend contours (different $\mathbf{g}_{hkl}$) and also for different sample areas, given by the variation in the width of bend contours due to the sample curvature as well as complex diffraction contrast in some areas of the samples. We estimated the variance based on an assumption that symmetrical bend contours (symmetry-equivalent $\mathbf{g}_{hkl}$) are expected in the model to have a single value, i.e. a mean bend contour displacement for symmetrical bend contours). A minimum of 3 $\mathbf{g}_{hkl}$ and corresponding bend contours would in principle allow an estimate of the variance for a particular $\mathbf{g}_{hkl}$. However, for the data presented here, few of the datasets show 3 or 4 measured for symmetry-equivalent $\mathbf{g}_{hkl}$ while the majority of datasets show only one or no available sets of symmetrical bend contour displacements. As a result, we evaluated the uncertainties across all datasets with the aim of estimating a global variance $\sigma^2_{global}$ for the general measurement of bend contour displacement in SED data. The variance for each symmetrical $\mathbf{g}_{hkl}$ follows:

$$\sigma^2 = \frac{1}{N}\sum_{i=1}^{N}(y_i - \bar{y})^2 \tag{S-14}$$

and the global variance is the average value of all the variances estimated from available symmetrical $\mathbf{g}_{hkl}$ across the datasets. Using this approach, we estimated the global variance $\sigma^2_{global}$ to be 71.2. This global variance corresponds to a global standard deviation $\sigma_{global}$ of 8.4. In turn, we used this global variance estimate for calculating $\chi^2_{red}$. We recognised that this estimate is imperfect due to differences in scatter of measurements for different bend contours within a SED dataset and between SED datasets, but we believe the $\chi^2_{red}$ nevertheless provides a useful summary statistic to evaluate the overall fitting approach and to compare fits with different number of parameters where necessary due to poorly constrained fitting parameters (high uncertainties in fitting parameters).

*Evaluation of uncertainties in fitting parameters*

The uncertainties in the fitting parameters in the simplified function f(φ) (equation 1) were calculated from the diagonals of the estimated covariance matrix of these parameters after fitting the data with the curve_fit function. The standard errors on the parameters were calculated by taking the square root of the diagonals of the covariance matrix. We also calculated the confidence interval for the fitting parameters using lmfit() function to determine how tightly the determined values of the fitting parameters are after fitting. In addition, a two-dimensional (2D) plot of the confidence region for each pair of fitting parameters was used to evaluate the dependence between the fitting parameters. For some datasets, fitting to equation 1 resulted in estimated standard errors for the fitting parameters larger than the fitting parameter magnitudes, suggesting these parameters were not well-constrained in the fitting to equation 1. We note that phenomenologically, there may be inherent ambiguity in constraining the parameters given limited data or where the available data can be fitted well with a range of compensating parameter settings. This off-setting property is particularly possible for the parameters *A* and *B* in equation 1 which both contribute to the scaling of the 'lobes' in the polar plots. In these cases, we have accordingly reduced the number of fitting parameters to avoid physically unreasonable fitting results.

In summary, our fitting evaluation procedure followed the workflow outlined below:

i. Carry out an initial fit to equation 1 $f(\varphi) = A \arctan(B \cos^2(\varphi - C))$
ii. Evaluate the suitability of equation 1 for the dataset:
  a. Evaluate the standard errors in the fitting parameters from the estimated covariance matrix.
  b. Evaluate confidence intervals and contour plot using Lmfit-1.2.0[2], a Python package for non-Linear least-squares minimization and curve-fitting.
  c. Analyse the residuals: A good model fit will exhibit residuals distributed symmetrically about zero. A poor model fit may exhibit multi-modal residual distributions or may exhibit a distribution centred away from zero.
  d. Calculate summary statistics ($R^2$, $\chi^2_{red}$)
iii. If equation 1 fitting is unsuitable, reduce the number of fitting parameters by adopting instead the fitting function (small-angle approximation to equation 1):

$$f(\varphi) = AB \cos^2(\varphi - C) = A_1 \cos^2(\varphi - C) \qquad \text{(S-15)}$$

iv. Evaluate the fit to equation S-15, including standard errors in parameters, Lmfit confidence intervals and contour plots, residuals, and summary statistics ($R^2$, $\chi^2_{red}$).

*Illustrative examples*

We illustrate this workflow in two example datasets: One that exhibits the properties of a well-constrained fit to equation 1 and a second that exhibits poorly constrained fitting to equation 1 and required fitting to equation S-15 instead.

The dataset depicted in Fig. 2c for p-terphenyl offers an example of a well-constrained fit to equation 1. Fitting bend contour displacements to equation 1 $f(\varphi) = A \arctan(B \cos^2(\varphi - C))$ gave the parameters $A$ = 144 ± 10 (nm); $B$ = 1.6 ± 0.2 (ratio); $C$ = 2.94 ± 0.01 (rad). The confidence interval calculated for these fitting parameters show:

|            | 95.45% | 68.27% | _BEST_ | 68.27% | 95.45% |
|------------|--------|--------|--------|--------|--------|
| $A$ (nm):  | -18    | -10    | 144    | +12    | +28    |
| $B$ (ratio):| -0.4  | -0.2   | 1.6    | +0.2   | +0.4   |
| $C$ (rad): | -0.03  | -0.01  | 2.94   | +0.01  | +0.03  |

In this case, the standard error calculated from the estimated covariance matrix closely resemble a 1σ range in the confidence intervals (68.3%). For the *B* and *C* parameters, the uncertainties are approximately linear when going from 1σ (68.3% confidence) to 2σ (95.0% confidence), and these values are fairly symmetric around the best fit values for fitting parameter B and C. There is an asymmetry distribution in the uncertainties of the fitting parameter A. The contour plots of confidence region, i.e. the maps of probability, for pairs of fitting parameters are shown in Supplementary Figure 4a to visualize the distribution of uncertainty and the correlations between the fitting parameters. These contour plots exhibit isotropic distributions for *A/C* and *B/C*, but a highly elliptical distribution for the *A/B* parameters, further elaborating the origin of the asymmetric confidence interval. Nevertheless, there remains a well-defined optimal fitting region for the *A* and *B* parameters.

The histogram of the residuals for fitting Fig. 2c with equation 1 is also plotted and compared to the residuals fitting with equation S-15 to further probe the relative goodness-of-fit for the two models (Supplementary Figure 4b). The residuals plots show a distribution centred at zero for equation 1, and show a bimodal, split distribution at negative and positive error for fitting equation S-15. The summary statistics ($R^2$, $\chi^2_{red}$) likewise confirm these conclusions, with values (0.97, 0.96) for fitting to equation 1 and (0.93, 2.35) for fitting to equation S-15.

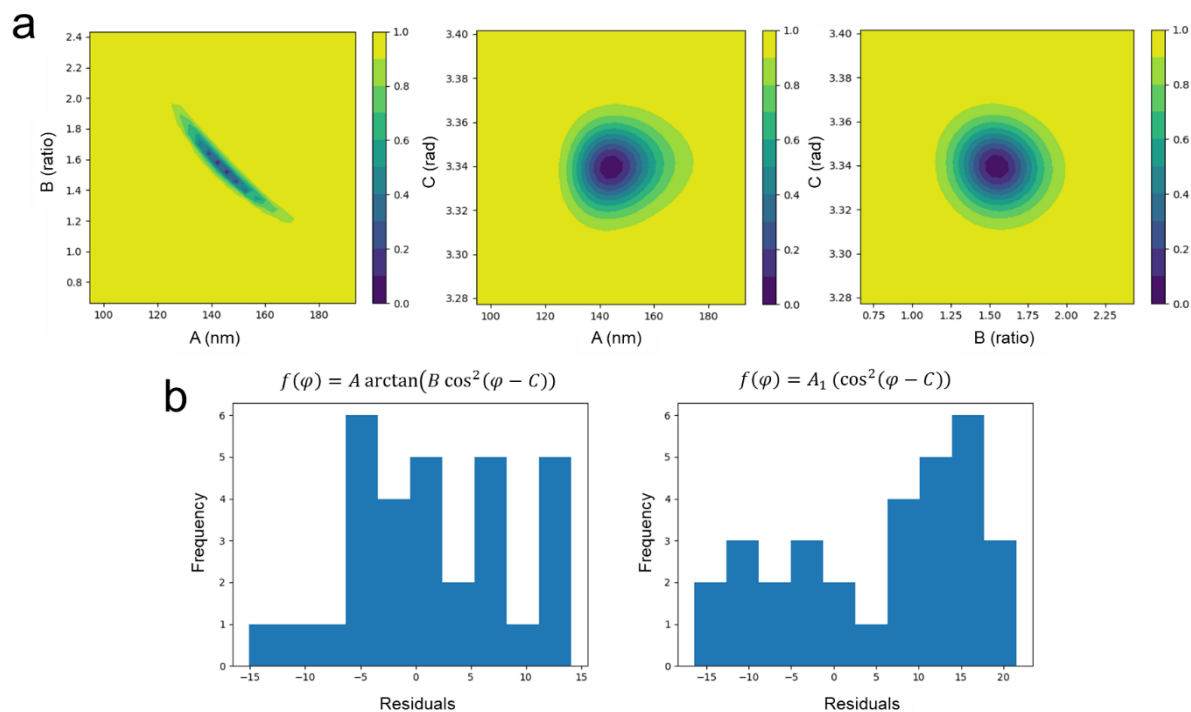

**Supplementary Figure 4. Evaluation of fitting procedure applied to measurements in Fig. 2c. a,** Contour plots of confidence region for each pair of fitting parameters showing the distribution of the uncertainties and the correlations between the fitting parameters: (left) correlation between A and B parameters, (middle) correlation between A and C parameters, and (right) correlation between B and C parameters. **b**, Histograms showing the distribution of the residuals for the two models used for the fitting.

Fig. 3d (anthracene) presents an example of a fit poorly constrained for fitting to equation 1 which required additional steps to finalise the fitting procedure. An initial fit to equation 1 gave the fitting parameters $A$ = 166800 ± 2e+10 (nm); $B$ = 0.0 ± 265.6 (ratio); and $C$ = 0.69 ± 0.03 (rad). The confidence intervals calculated for these fitting parameters were determined as:

|          | 95.45%   | 68.27%   | _BEST_  | 68.27% | 95.45% |
|----------|----------|----------|---------|--------|--------|
| A (nm):  | -166568  | -166360  | 166800  | +inf   | +inf   |
| B (ratio): | -inf   | -inf     | 0.0     | +inf   | +inf   |
| C (rad): | -0.06    | -0.03    | 0.69    | +0.03  | +0.06  |

These fitting results exhibited standard errors much larger than the *A* and *B* fitting parameter magnitudes. The fitting parameter *C* is well constrained, likely because this parameter defines the orientation of the 'lobes' in the polar plot. This orientation parameter appears to be well-separated from the scaling parameters *A* and *B*. The calculated confidence intervals for the *A* and *B* parameters are very asymmetric and approach negative and positive infinity (-inf and +inf). In other words, there is insufficient data to unambiguously determine the fitting parameters *A* and *B* in the model. Effectively, many sets of *A* and *B* parameters will generate curves that fit the data equally well. The contour plots of confidence region (Supplementary Figure 5a) show no specific distribution for *A* and *B* which highlights that they are not well constrained.

In order to evaluate the effect of the initial guess on these results, we refined the fitting process by fitting fewer parameters. First, we fixed the *B* parameter and allowed *A* and *C* to vary in the fit and then used these values as initial guess values to fit all three parameters. Following the first step by fixing B = 0.5, we obtained the parameters *A* = 1005 ± 54 (nm) and C = 0.69 ± 0.02 (rad). These fitting results show reasonable estimates of the standard errors for *A* and *C*. After the second, three-parameter fit, the parameters became *A* = 169205 ± 2e+10 (nm); *B* = 0.0 ± 269.3; and *C* = 0.69 ± 0.03. These fitting results illustrate the instability in the fit, persistent for stepwise parameter fitting.

Consequently, we turned to a reduction in the number of parameter by simplification of the model. Parameters *A* and *B* together set the length and width of the 'lobes' in the polar plot. For small B or small values of $\cos^2(\varphi - C)$ the curve (equation 1) approaches the small-angle approximation: $f(\varphi) = A_1 \cos^2(\varphi - C)$ (equation S-15) which combines *A* and *B* into a single parameter *A₁*. Notably, $\cos^2(\varphi - C)$ will be small near $\mathbf{g}_{hkl} \cdot \mathbf{B} = 0$, providing suitable justification for fitting equation S-15 when measurements are only available near the zero-displacement condition for visible bend contours. More generally, incomplete sampling of the width and length of the 'lobes' by experimental measurements will offer poor constraints on *A* and *B*, and its replacement with a robust 2-parameter fit to equation S-15 retains the targeted form for identifying the $\mathbf{g}_{hkl} \cdot \mathbf{B} = 0$ as defined by $\cos^2(\varphi - C)$.

In Fig. 3d, by fitting equation S-15 we obtained *A₁* = 493 ± 27 and *C* = 0.69 ± 0.02. The calculated confidence intervals were determined as:

|  | 95.4% | 68.3% | _BEST_ | 68.3% | 95.5% |
|---|---|---|---|---|---|
| *A₁* (nm): | -62 | -28 | 493 | +28 | +62 |
| *C* (rad): | -0.06 | -0.03 | 0.69 | +0.02 | +0.05 |

The standard errors on the parameters are significantly smaller than the parameter values in this case, suggesting well-constrained parameter fits. The calculated confidence intervals also show approximately linear changes from 1σ (68.3% confidence) to 2σ (95.5% confidence), and these values exhibit symmetry around the best fit values for *A₁* and *C*. The contour plots of confidence region *A₁*/*C* (Supplementary Figure 5b) exhibits an approximately isotropic distribution with no features indicating significant correlation between *A₁* and *C*. The residuals plots obtained for both fitting models are indistinguishable (Supplementary Figure 5c), highlighting the correspondence between equation 1 and its small-angle approximation (equation S-15) for the available measurements in this example. The summary statistics ($R^2$, $\chi^2_{red}$) were (0.92, 13.55) for fitting to equation 1 and (0.93, 12.04) for fitting to equation S-15, indicating similar summary statistics for both models albeit slightly improved for fitting to equation S-15. The $\chi^2_{red}$ in this final fitting (equation S-15) is significantly greater than 1, indicating the residuals exceed the expected variance estimated by $\sigma^2_{global}$. The global variance may not correspond to the variance in the bend contour displacements in this dataset or across the available $\mathbf{g}_{hkl}$ for this dataset. Reviewed together, the equation S-15 fit offers a best fit with constrained parameters and unambiguously determines the orientation of the 'lobes' and therefore the $\mathbf{g}_{hkl} \cdot \mathbf{B} = 0$ condition. We note that across all datasets, the *C* parameter captures the key information for estimating the Burgers vector as it fits the orientation (azimuthal angle) of the 'lobes' of the polar plot.

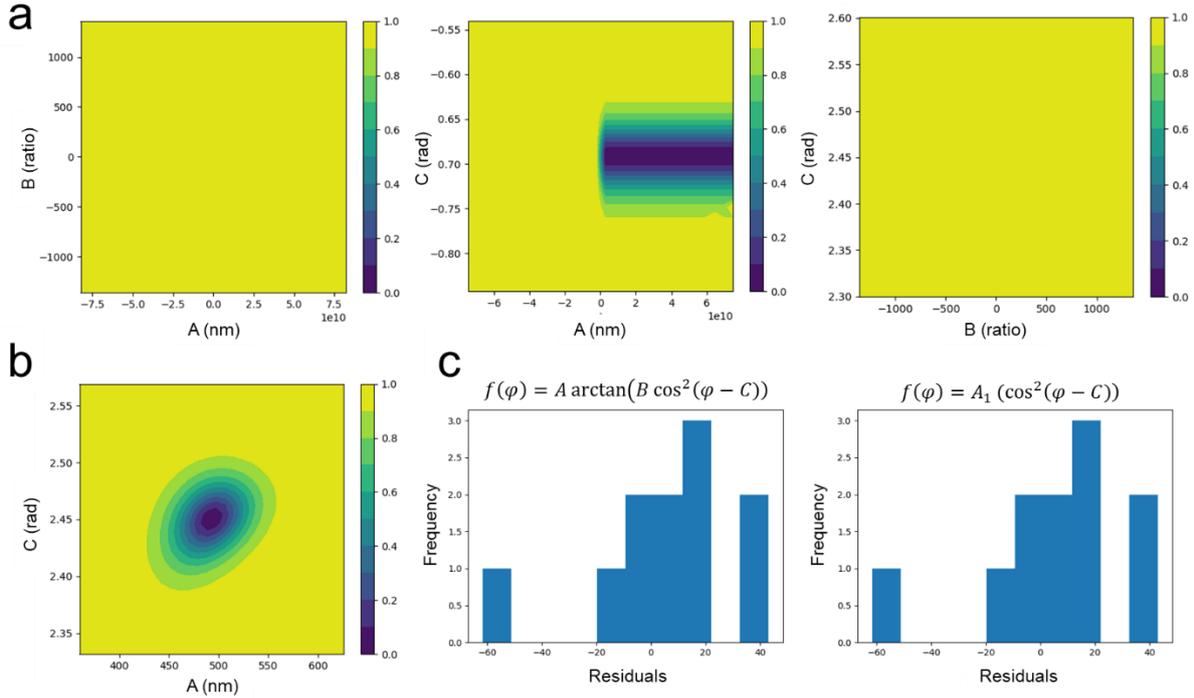

**Supplementary Figure 5. Evaluation of fitting procedure applied to measurements in Fig. 3d. a,** Contour plots of confidence region for each pair of fitting parameters using $f(\varphi) = A \arctan(B \cos^2(\varphi - C))$ as a fitting model: (left) correlation between A and B parameters, (middle) correlation between A and C parameters, and (right) correlation between B and C parameters. **b,** Contour plots of confidence region for A and C parameters using $f(\varphi) = A_1 \cos^2(\varphi - C)$ as a fitting model. **c,** Histograms showing the distribution of the residuals for two models used for the fitting.

**Supplementary Table 2. Summary of fitting coefficients.** The fitting coefficients are given with uncertainties reported as standard errors determined from the diagonals of the estimated covariance matrix. The coefficient of determination $R^2$, the number of data points $N$, degrees of freedom $\nu$ and $\chi^2_{red}$ are also reported. Rows containing fitting results for the parameter $B$ report fitting to equation 1 ($A$, $B$, $C$ parameters). Rows containing no $B$ parameter fit show fitting to equation S-15 ($A_1$, $C$ parameters). Rows containing a further $D$ parameter used an additional form adapted for cases with two unresolved dislocation contributions (see Supplementary Fig. 18-19).

| Figure | $A$ or $A_1$ (nm) | $B$ (ratio) | $C$ (rad) | $D$ (nm) | $R^2$ | $N$ | $\nu$ | $\chi^2_{red}$ |
|---|---|---|---|---|---|---|---|---|
| Fig. 2c | 144 ± 9.9 | 1.6 ± 0.2 | 2.94 ± 0.01 | - | 0.97 | 31 | 28 | 0.96 |
| Fig. 3d | 493 ± 26.6 | - | 0.69 ± 0.02 | - | 0.93 | 11 | 9 | 12.0 |
| Fig. 4d | 115 ± 7.8 | 2.6 ± 0.5 | 2.49 ± 0.03 | - | 0.98 | 14 | 11 | 0.81 |
| Fig. 4h | 144 ± 5.6 | - | 1.95 ± 0.02 | - | 0.97 | 21 | 19 | 0.94 |
| Supp. Fig. 6-2 | 148 ± 2.2 | - | 2.96 ± 0.01 | - | 0.98 | 38 | 36 | 0.81 |
| Supp. Fig. 6-3 | 111 ± 20.4 | 1.3 ± 0.4 | 2.91 ± 0.02 | - | 0.94 | 23 | 20 | 1.15 |
| Supp. Fig. 7b-1 | 67 ± 9.3 | - | 2.74 ± 0.05 | - | 0.85 | 15 | 13 | 0.39 |
| Supp. Fig. 7b-2 | 115 ± 36.7 | 1.5 ± 0.7 | 2.78 ± 0.03 | - | 0.95 | 13 | 10 | 0.74 |
| Supp. Fig. 7b-3 | 110 ± 5.3 | - | 2.86 ± 0.02 | - | 0.98 | 16 | 14 | 0.15 |
| Supp. Fig. 8c-2 | 117 ± 9.8 | - | 3.08 ± 0.05 | - | 0.92 | 14 | 12 | 1.03 |
| Supp. Fig. 8c-3 | 109 ± 3.1 | - | 2.96 ± 0.02 | - | 0.97 | 15 | 13 | 0.37 |
| Supp. Fig. 12a-2 | 108 ± 8.1 | - | 0.34 ± 0.04 | - | 0.85 | 11 | 9 | 1.89 |
| Supp. Fig. 13b-1 | 517 ± 7.3 | - | 0.75 ± 0.02 | - | 0.96 | 11 | 9 | 6.00 |
| Supp. Fig. 13b-2 | 198 ± 16.3 | - | 1.69 ± 0.03 | - | 0.94 | 11 | 9 | 1.47 |
| Supp. Fig. 18c | 155 ± 9.1 | - | 3.13 ± 0.03 | 56.73 ± 3.67 | 0.96 | 15 | 12 | 1.53 |
| Supp. Fig. 19b | 903 ± 95.2 | - | 0.79 ± 0.03 | 331.75 ± 19.6 | 0.90 | 14 | 11 | 27.6 |

**Supplementary Note 3: Displacement of bend contours at parallel g vectors**

The proposed model describes the displacement of a bend contour arising from the rotation of $\mathbf{g}_{hkl}$ and the corresponding tilt of lattice planes associated to $\mathbf{g}_{hkl}$ at a dislocation core. Thus, the displacement of the bend contours constructed from a set of parallel $\mathbf{g}_{hkl}$, i.e. parallel sets of planes, should be equal as the tilt for these sets of parallel planes is the same. As a demonstration of this geometric principle, Supplementary Figure 6a shows approximately equal displacements of the $\mathbf{g}_{210}$ and $\mathbf{g}_{420}$ bend contours. The width of the bend contour for $\mathbf{g}_{420}$ is smaller than the width of the bend contour for $\mathbf{g}_{210}$. Additionally, a node appears at the split of bend contours for $\mathbf{g}_{420}$. Based on $\mathbf{B} = [010]$ assigned for dislocation-1 (Fig. 2), we note that $\mathbf{B} = [010]$ is aligned with the dislocation line $\mathbf{u}$ for dislocation-1 in the local area where dislocation-1 cuts through the $\mathbf{g}_{420}$ bend contour. This observation points to a predominantly screw dislocation character at this position in the field of view, matching expectations for $\mathbf{g}_{hkl} \cdot \mathbf{B}_{screw} = n$ for integer $n$ with $n - 1$ nodes[3].

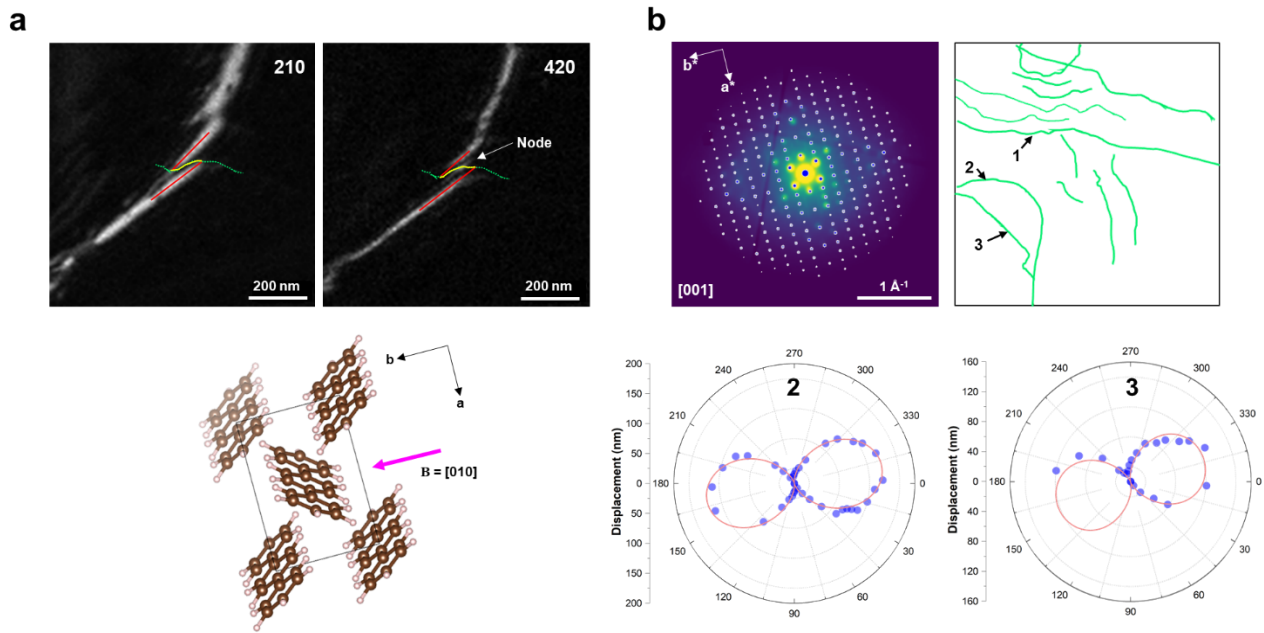

**Supplementary Figure 6. Further analysis of dislocations in p-terphenyl in Fig. 2. a,** Comparison between the displacements of the bend contours corresponding to parallel sets of planes, i.e. $\mathbf{g}_{210}$ and $\mathbf{g}_{420}$, at the dislocation. The node appears at the splitting of the bend contour for $\mathbf{g}_{420}$ where dislocation-1 is almost parallel to $\mathbf{B} = [010]$. **b,** Demonstration of this analysis approach on dislocation-2 and dislocation-3: (upper left) diffraction patterns (with on-zone simulated pattern overlaid in white) taken from the whole field of view, and (upper right) the mapping of the dislocations presented in the field of view. The polar plot of the bend contour displacements via azimuthal angle φ in diffraction space for dislocations-2 and -3 (bottom left and bottom right, respectively) show good fit to the simplified version of Equation S-3, similar to dislocation-1. The continuous bend contours, which are not affected by dislocations 2 and 3, are indexed as 200 and $\bar{2}00$ corresponding to the Burgers vector $\mathbf{B} = [010]$. Modelling of the $\mathbf{B}$ direction in p-terphenyl unit cell viewing along [001] with adjusted in-plane rotation to match the orientation of the diffraction pattern from the experimental dataset showing the mixed-type dislocation for the two dislocations.

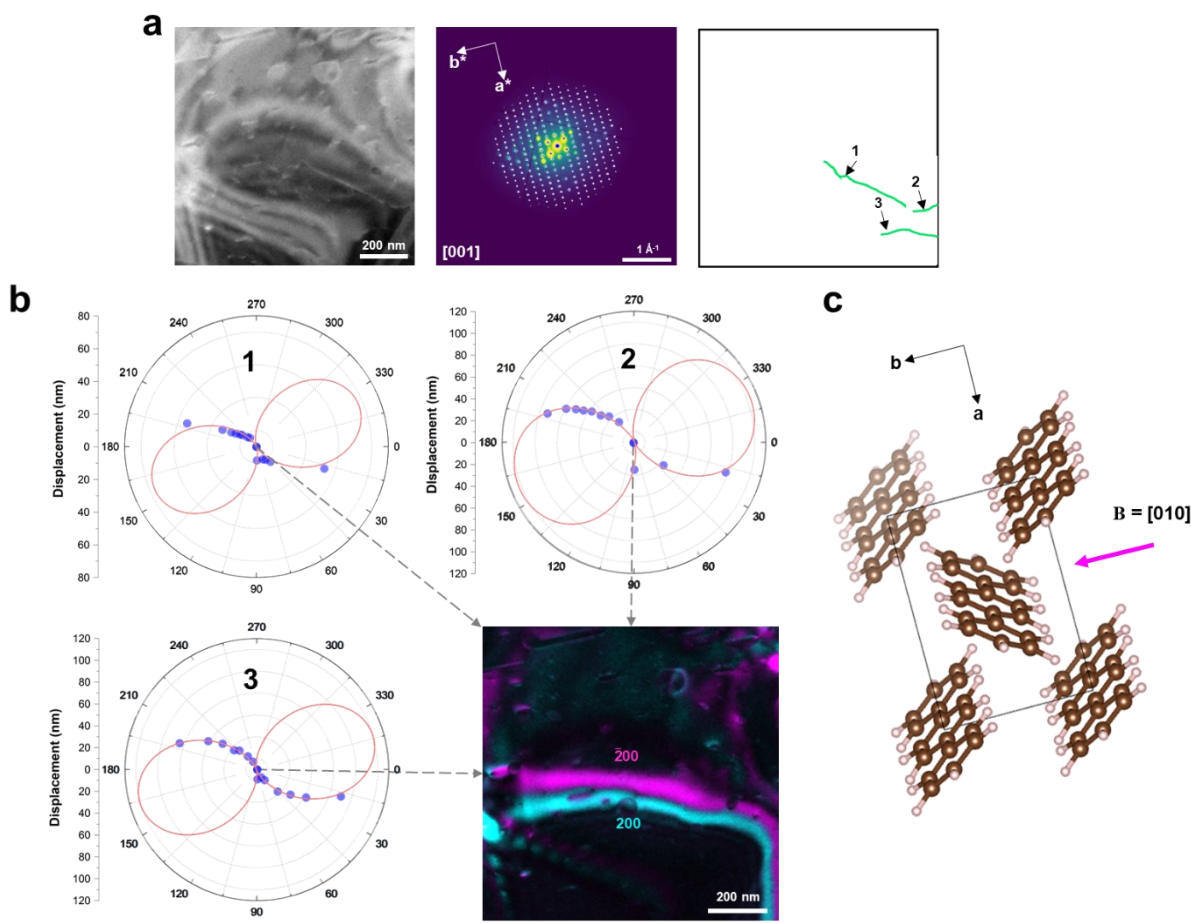

**Supplementary Figure 7. Demonstration of the analysis approach on dislocations having only a few measurements of bend contour displacements. a,** (Left) Annular dark field (ADF) image showing the pole bend contours and several basal and non-basal dislocations on the p-terphenyl film, (middle) corresponding diffraction pattern (with on-zone simulated pattern overlaid in white) indicating the slightly off zone axes of [001] which is in agreement with the pole bend contours slightly outside the field of view in the ADF image, and (right) in-plane dislocations are identified and presented by the green lines showing three small line dislocations. **b,** Polar plot of the bend contour displacement via azimuthal angle φ in diffraction space showing the non-complete plot for all three dislocations due to the limited field of view (1000x1000 nm) that does not allow to construct the observable bend contours outside the field of view. The fitting function following the simplified version of Equation S-3 is used to estimate the complete polar plot in these cases. The VDF image of the continuous bend contours can be constructed for the three dislocations showing the 200 and $\bar{2}00$ as the common continuous bend contours. The Burgers vector of all three dislocations, thus, follows **B** = [010]. **c,** Modelling of the **B** direction in p-terphenyl unit cell viewing along [001] direction with adjusted in-plane rotation to match the orientation of the diffraction pattern from experimental data set showing mixed-type for dislocation-1, pure screw for dislocation-2, and screw to mixed-type for dislocation-3.

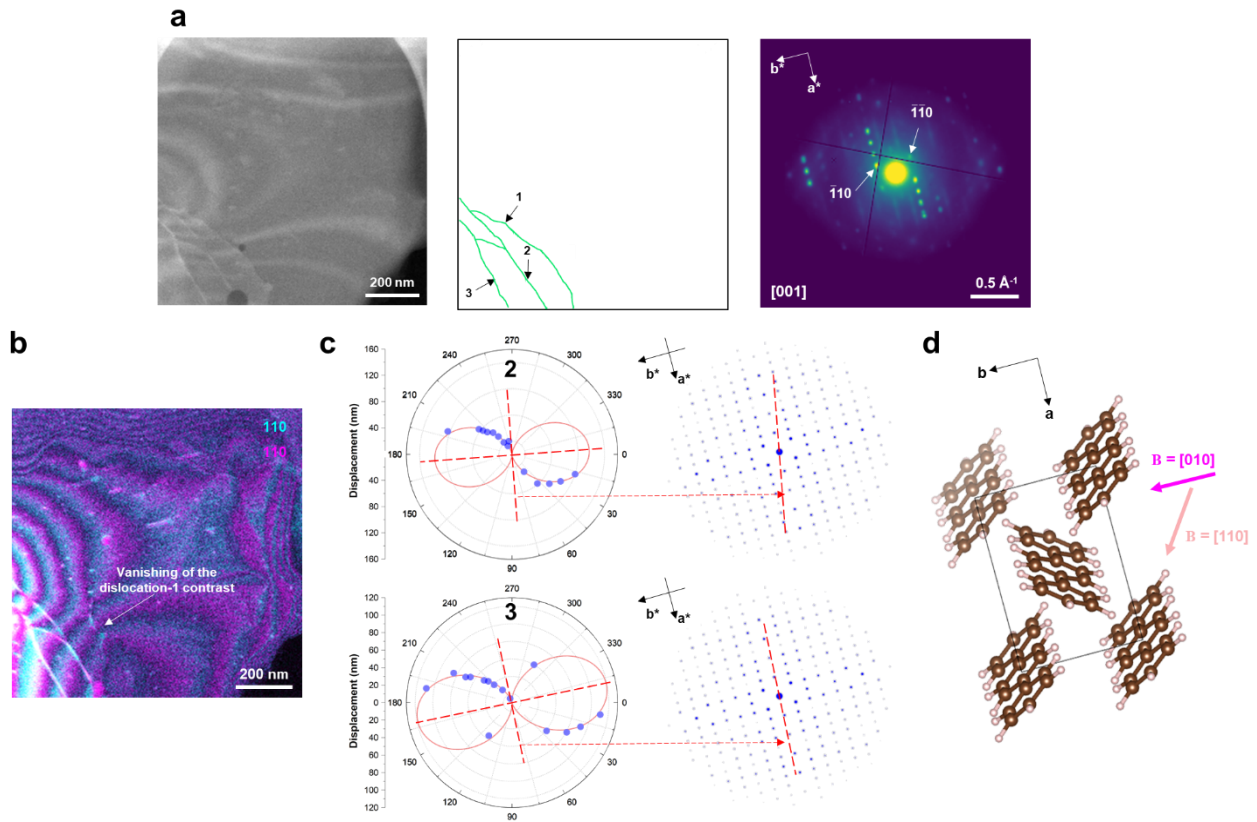

**Supplementary Figure 8. Demonstration of the analysis approach for a p-terphenyl film tilted approximately 3.6° from the [001] zone axis. a,** (Left) ADF image (1000x1000 nm) showing several bend contours crossing a dislocation network comprising of three major dislocation lines, (middle) in-plane dislocations are presented by the green lines and numbered, and (right) corresponding diffraction pattern indicating the significant tilting away from the [001] zone axes as seen by the absence of the majority of diffraction spots/discs. **b,** VDF image constructed from the diffraction vector $g_{\bar{1}10}$ and $g_{1\bar{1}0}$ showing the invisibility of dislocation-1 while dislocation-2 and -3 show an enhanced contrast. This suggests the invisibility criterion condition for dislocation-1 at these diffraction vectors, thus, the Burgers vector **B** = [110] can be assigned. **c,** Polar plot of the bend contour displacement via azimuthal angle φ constructed for dislocation-2 and -3 (with on-zone simulated pattern in blue added next to the plots) showing the non-complete plot and the absence of the diffraction vectors at the invisibility criterion condition for these dislocations. The fitting function following the simplified version of Equation S-3 estimates the complete polar plot that shows the invisibility criterion condition for dislocation-3 at $g_{200}$ and $g_{\bar{2}00}$. The Burgers vector of dislocation-3 is therefore **B** = [010]. However, the fitting function estimates the invisibility criterion condition for dislocation-2 at $g_{\bar{8}10}$ and $g_{810}$ which gives the Burgers vector $[u_B v_B 0]$ = [1$\bar{8}$0] which is 10.2° away from the common **B** = [010]. This deviation of dislocation-2 Burgers vector may either arise from the residual uncertainties in bend contour measurements or significant tilting of the sample from the [001] zone axes. **d,** Modelling of the two **B** directions in p-terphenyl unit cell viewing along [001] direction with adjusted in-plane rotation to match the orientation of the diffraction pattern from experimental data set showing edge-type for dislocation-1 and mixed-type for dislocation-2 and dislocation-3.

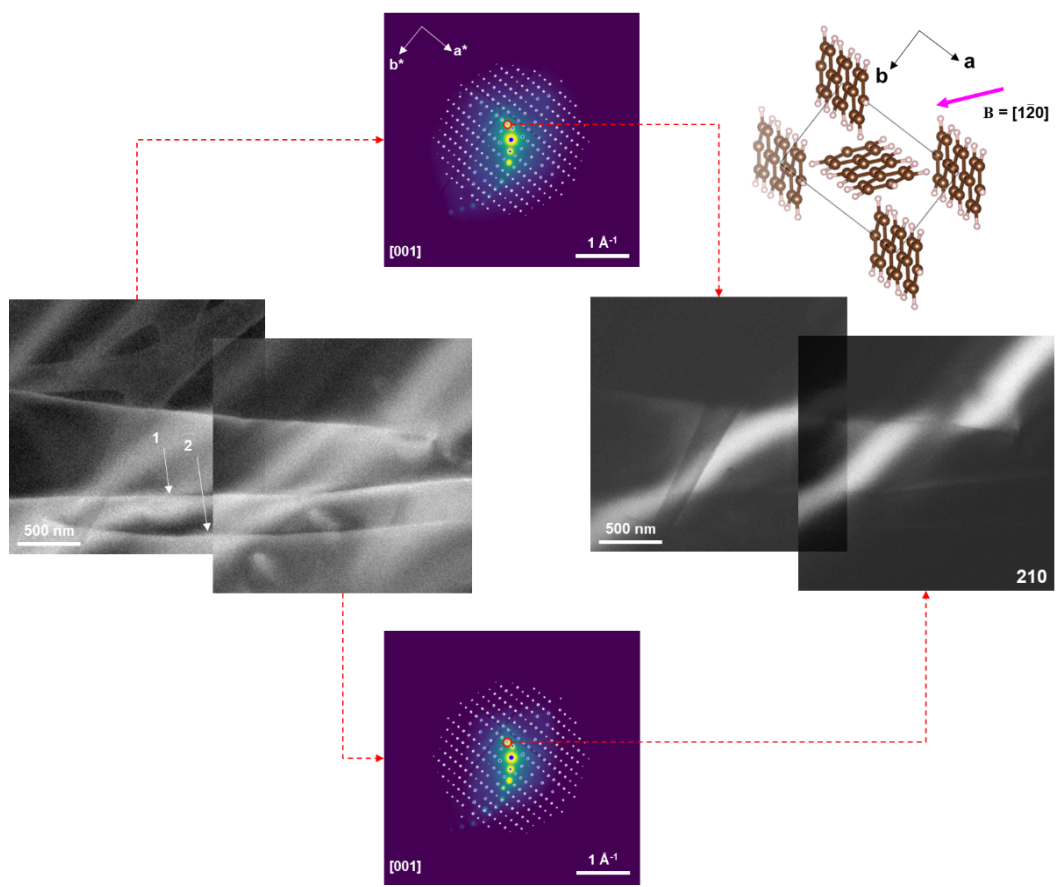

**Supplementary Figure 9. SED measurement at two adjacent areas on the p-terphenyl film.** ADF images showing the overlapping region between two SED datasets which can be used to merge two images. Corresponding diffraction patterns (with on-zone simulated pattern overlaid in white) show significant tilting of the film away from the [001] zone axes. The VDF image constructed at the same $\mathbf{g}_{210}$ for these two datasets showing the shift of the 210 bend contour between two SED measurements. It suggests the re-orientation of the film under the scanning of the electron beam. The 210 bend contour shows no break on crossing dislocations-1 and -2 suggesting that these two dislocations have the same invisibility criterion condition at $\mathbf{g}_{210}$. The Burgers vector, thus, follows $\mathbf{B} = [1\bar{2}0]$. Modelling of the $\mathbf{B}$ direction (upper right) in p-terphenyl unit cell viewing along [001] direction with adjusted in-plane rotation to match the orientation of the diffraction pattern from experimental data set showing mixed-type for dislocation-1 and mixed-type for dislocation-1 and dislocation-2.

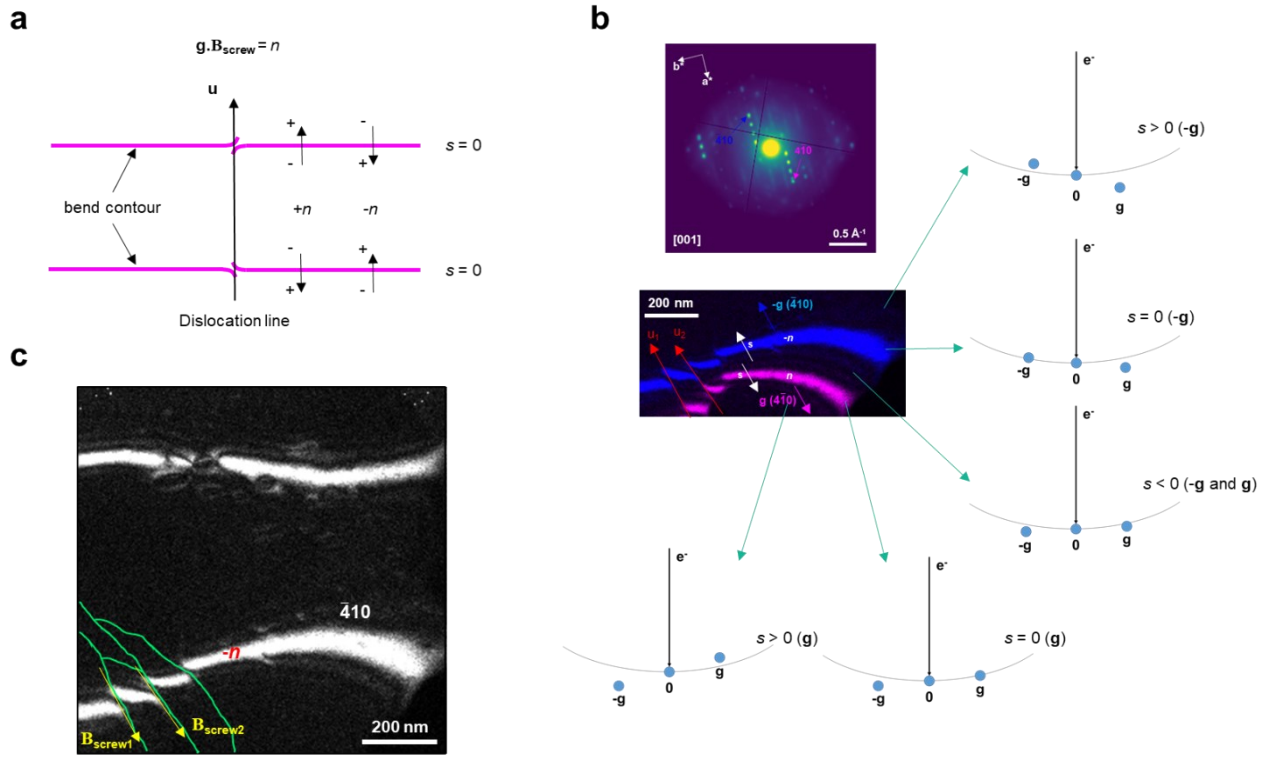

**Supplementary Figure 10. Relative handedness of screw components between individual dislocations. a,** Application of Cherns and Preston rules for the splitting and twisting of bend contours[4] showing the determination of the signs of n in $\mathbf{g}\cdot\mathbf{B}_{screw} = n$ condition where $\mathbf{B}_{screw}$ is the screw component of the Burgers vector $\mathbf{B}$. A 'twisting' of the bend contour highlights a screw character of dislocations which is due to the continuous change of the excitation error from negative to positive or vice versa of the diffracting planes near the dislocation on approaching the dislocation core (specific for screw-type components)[3]. **b,** Determination of excitation error ($s$) on either side of the bend contour using a pair of the $\mathbf{g}_{hkl}$ and $-\mathbf{g}_{hkl}$ bend contours from the VDF image. **c,** Analysis of the relative handedness of screw components between dislocation-2 and dislocation-3 in Supplementary Figure 6 using the twisting of the bend contour $\bar{4}01$. The direction of the dislocations are kept similar and depicted as $u_1$ and $u_2$. By using the approaches in **a** and **b**, we can determine the condition in $\mathbf{g}\cdot\mathbf{B}_{screw} = -n$ for both dislocation-2 and -3 for the same $\mathbf{g}_{\bar{4}10}$, thus, indicating the same handedness of the screw components of the two dislocations.

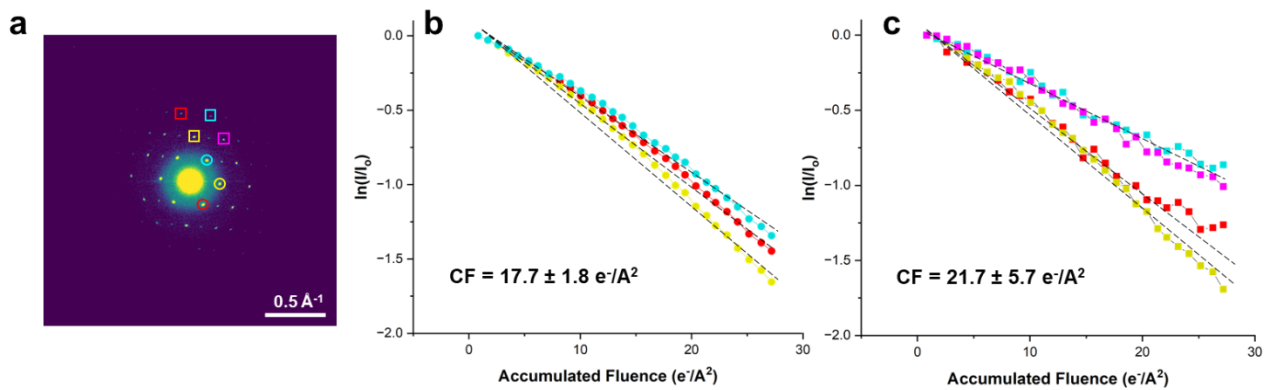

**Supplementary Figure 11. Loss of Bragg diffraction intensity with cumulative electron fluence (beam-damage series) for an anthracene thin film viewed along [101]. a,** Selected Area Electron Diffraction Pattern (SAED) taken at the beginning of the experiment ($t_o$) showing the marked diffraction spots for CF analysis. **b**, Decay profile of the lattice planes having *d*-spacing > 4Å. **c**, Decay profile of the lattice planes having *d*-spacing < 4Å. The decay profiles showed an approximately linear response with an adjusted determination coefficient ($R^2$) > 0.9. The CF is extracted from the gradient of the linear fit (see Supplementary Note 4).

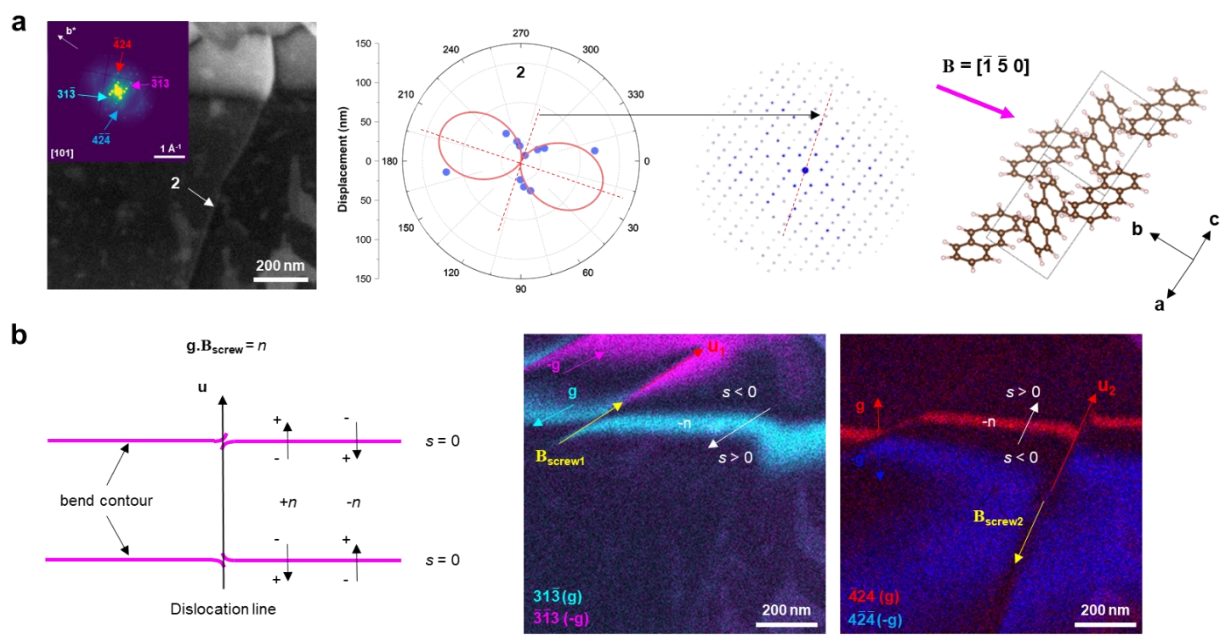

**Supplementary Figure 12. Analysis of a second dislocation in the anthracene film in Fig. 3. a,** (Left) ADF-STEM image with average electron diffraction pattern inset, (middle) polar plot of bend contour breaks at a dislocation line and accompanying fit displayed on a simulated diffraction pattern, and (right) visualisation of the crystal structure with determined Burgers vector shown. The incomplete polar plot of the bend contour displacement on crossing dislocation-2 via azimuthal angle φ shows good fit to the simplified version of Equation S-3, similar to dislocation-1. There is an absence of the diffraction vectors at the invisibility criterion condition for this dislocation and the fitting function estimates the complete polar plot showing such condition at ~ $g_{5\bar{1}\bar{5}}$ and $g_{\bar{5}15}$ for dislocation-2. The Burgers vector of this dislocation, determined using equation S-7 for the monoclinic anthracene crystal where $a \cdot b \neq 0$, follows $B = [u_B, v_B, w_B]$ with $u_B$ = -0.16, $v_B$ = -1.00, $w_B$ = 0.04 which is approximately along $[u_B v_B 0] = [\bar{1}\bar{5}0]$ (3.6° from perpendicular to [101] and 3.6° from extracted **B**, 1.5 standard errors for a standard error of 0.04 radians). Modelling of the **B** direction in anthracene unit cell viewing along [101] direction with adjusted in-plane rotation to match the orientation of the diffraction pattern from the experimental data set showing mixed-type dislocation for dislocation-2 but with an edge component that is predominant. **b,** Relative handedness of screw components between dislocation-1 and dislocation-2 in anthracene film: (Left) application of Cherns and Preston rules for the splitting and twisting of bend contours, (middle) determination of excitation error (s) and $B_{screw}$ direction for dislocation-1, and (right) determination of excitation error (s) and $B_{screw}$ direction for dislocation-2. Following the determination of excitation error $s$ and the Cherns and Preston rule for $B_{screw}$ component, the condition of $g \cdot B_{screw1}$ = -$n$ is obtained for dislocation-1 with $g_{31\bar{3}}$. By applying the same procedure for dislocation-2, the condition $g \cdot B_{screw2}$ = -$n$ is also obtained with $g_{\bar{4}24}$. Comparing the direction of $B_{screw1}$ and $B_{screw2}$ shows the opposite handedness of screw components of the dislocation-1 and -2 in the anthracene film.

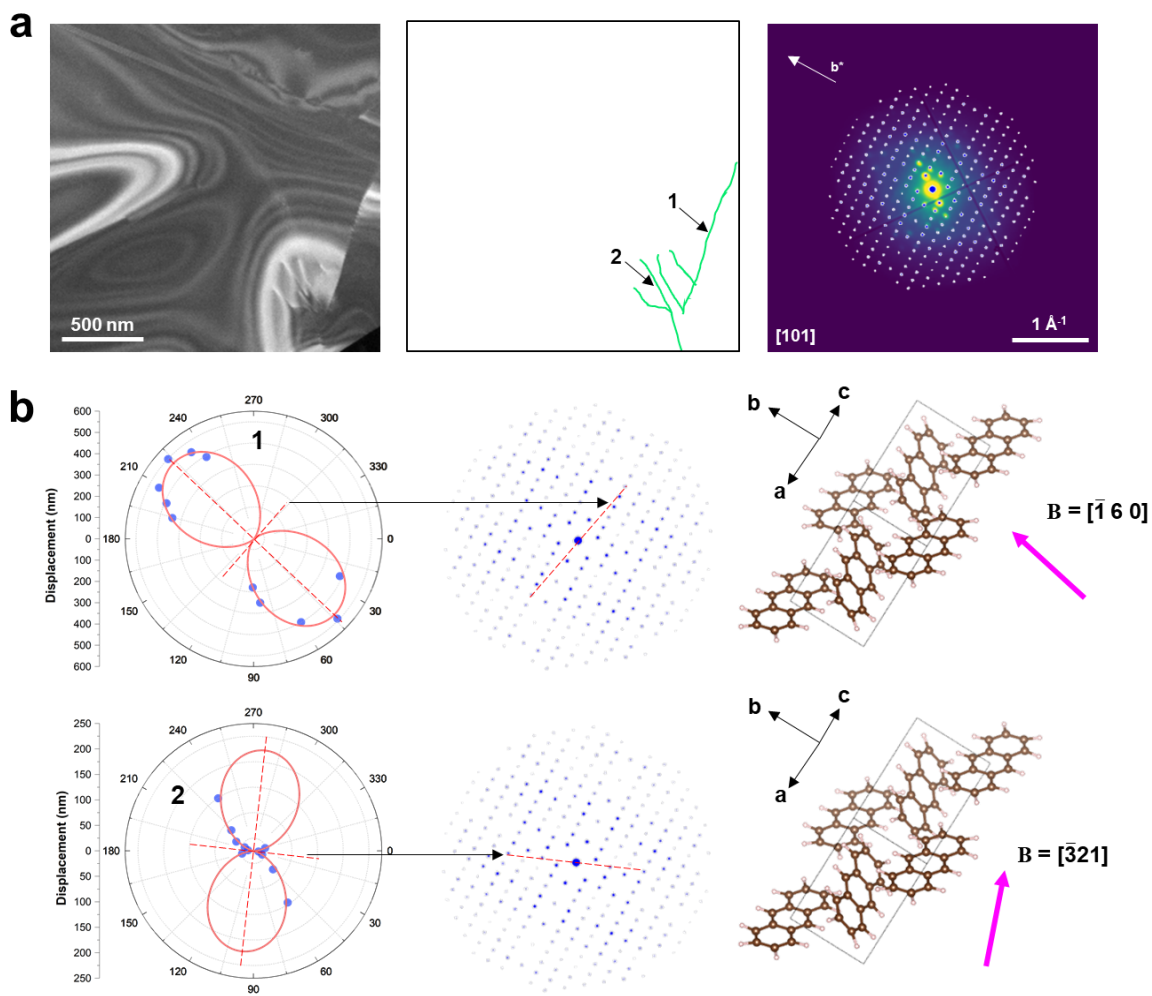

**Supplementary Figure 13. Application of the analysis approach on the anthracene film prepared by spin coating. a,** (Left) ADF image showing breaks in bend contours on crossing a dislocation network, (middle) in-plane dislocations are presented by the green lines and are numbered, and (right) corresponding diffraction pattern (with on-zone simulated pattern overlaid in white) indicating the significant tilting of the film away from the [101] zone axes as seen by the absence of the majority of diffraction spots/discs shown in the overlaid zone axis simulation (white spots). **b,** (Left) Polar plot of the bend contour displacement via azimuthal angle φ constructed for dislocation-1 and -2 showing an incomplete plot and the absence of the diffraction vectors at the invisibility criterion condition for these dislocations. The fitting function following the simplified version of Equation S-3 provides the complete polar plot that shows the invisibility criterion condition at $g_{61\bar{6}}$ and $g_{\bar{6}1\bar{6}}$ for dislocation-1 and $g_{12\bar{1}}$ and $g_{\bar{1}2\bar{1}}$ for dislocation-2 after indexation to the on-zone simulated diffraction pattern (middle). (Right) The Burgers vector for dislocation-1 follows $\mathbf{B} = [u_B, v_B, w_B]$ with $u_B = -0.13$, $v_B = 1.00$, $w_B = 0.04$ which is approximately along $[u_B v_B 0] = [\bar{1}60]$ (3.1° from perpendicular to [101] and 3.1° from extracted $\mathbf{B}$, 2.5 standard errors for a standard error of 0.02 radians). The Burgers vector for dislocation-2 follows $\mathbf{B} = [u_B, v_B, w_B]$ with $u_B = -0.13$, $v_B = -1.00$, $w_B = 0.04$ which is approximately along $[\bar{3}21]$ (2.4° from perpendicular to [101] and 2.4° from extracted $\mathbf{B}$, 1.3 standard errors for a standard error of 0.03 radians). This dislocation-2 Burgers vector has not been reported previously nor would appear to be readily expected; it may either arise from the method of preparation, proximity to other dislocations, or from limitations in the dataset due to residual uncertainties in bend contour measurements or sample orientation effects.

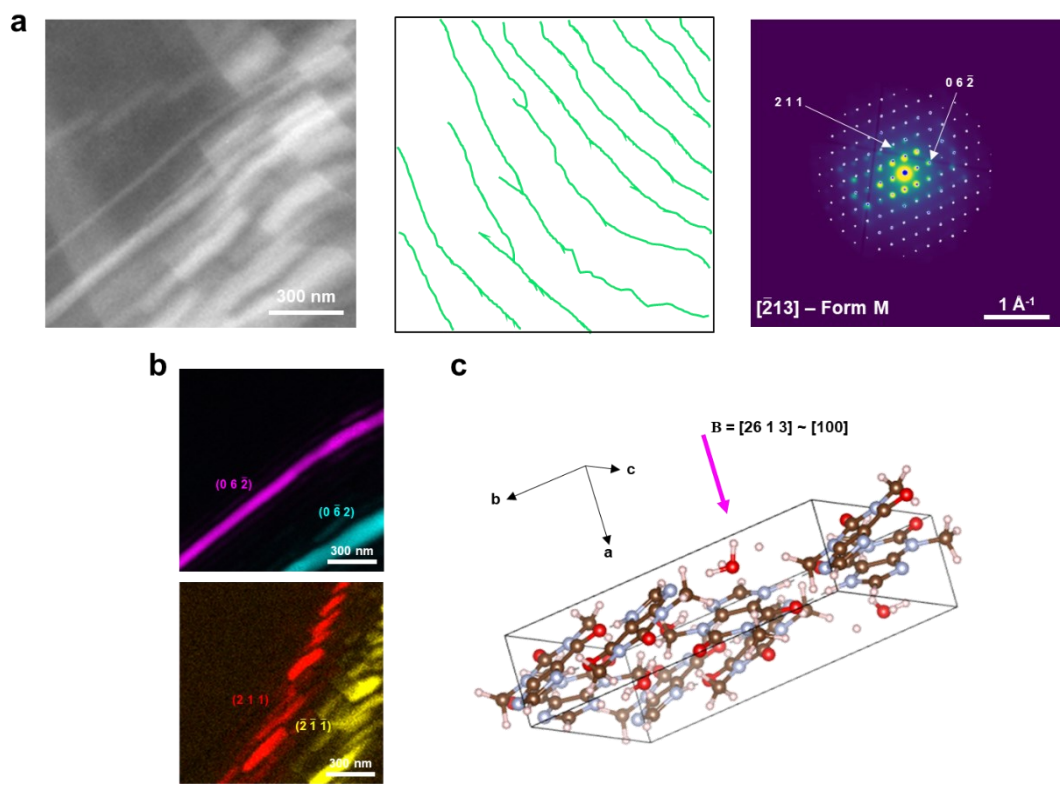

**Supplementary Figure 14. Analysis of dislocation network in theophylline monohydrate (Form M) prepared by the solvent evaporation from nitromethane:water solution. a,** (Left) ADF image showing several bend contours on crossing a large dislocation network across the field of view. Individual dislocations in the dislocation network are presented by the green lines. Corresponding diffraction pattern (with on-zone simulated pattern overlaid in white) showing the monohydrate theophylline film crystallised on the [$\bar{2}$13] zone axis. **b,** VDF images of the bend contour pairs at different **g.B** conditions. VDF image of the bend contour pair constructed from $g_{06\bar{2}}$ and $g_{0\bar{6}2}$ showing the continuous bend contours on crossing several dislocations which highlights the invisibility criterion condition at these two diffraction vectors. VDF image of the bend contour pair constructed from $g_{211}$ and $g_{\overline{211}}$ showing the break of bend contours at individual dislocations. The Burgers vector for individual dislocations in the dislocation network follows **B** = [$u_B$, $v_B$, $w_B$] with $u_B$ = 1.00, $v_B$ = 0.04, $w_B$ = 0.12 or **B** = [26 1 3]. This unlikely high index vector aligns closely with the projection of the [100] direction, and therefore the Burgers vector was assigned as **B** = [100]. **c,** Modelling of the **B** direction in theophylline form M unit cell viewing along [$\bar{2}$13] with adjusted in-plane rotation to match the orientation of the diffraction pattern from experimental data set.

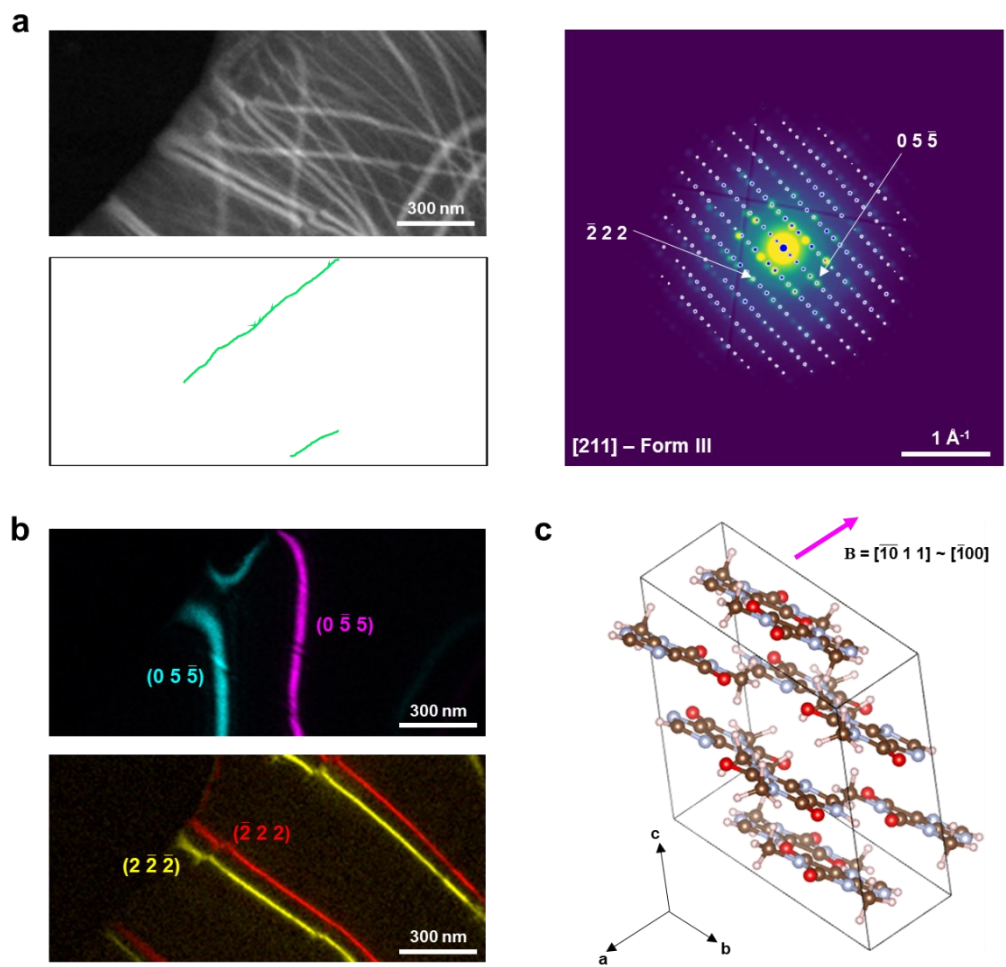

**Supplementary Figure 15. Analysis of dislocation network in metastable theophylline (Form IIIb) particle.** Data acquisition was carried out under liquid nitrogen cooling with the cryo-holder. **a,** ADF image showing bend contour network on crossing dislocations at some areas of the film. Individual dislocations are presented by green lines at which the bend contours show breaking in the ADF image. Corresponding diffraction pattern (with on-zone simulated pattern overlaid in white) showing theophylline From III film viewed along the [211] zone axis. **b,** VDF images of the bend contour pairs at different **g.B** conditions. VDF image of the bend contour pair constructed from $g_{05\bar{5}}$ and $g_{0\bar{5}5}$ showing the continuous bend contours on crossing dislocations which highlights the invisibility criterion condition at these two diffraction vectors. VDF image of the bend contour pair constructed from $g_{2\bar{2}\bar{2}}$ and $g_{\bar{2}22}$ showing the break of bend contours at individual dislocations. The Burgers vector for individual dislocations in the dislocation network follows **B** = [$u_B$, $v_B$, $w_B$] with $u_B$ = -1.00, $v_B$ = 0.10, $w_B$ = 0.10 or **B** = [$\bar{1}0$ 1 1]. This unlikely high index vector aligns closely with the projection of the [100] direction, and therefore the Burgers vector was assigned as **B** = [$\bar{1}$00]. **c,** Modelling of the **B** direction in theophylline form III unit cell viewed along [211] with adjusted in-plane rotation to match the orientation in the diffraction pattern from experimental data set.

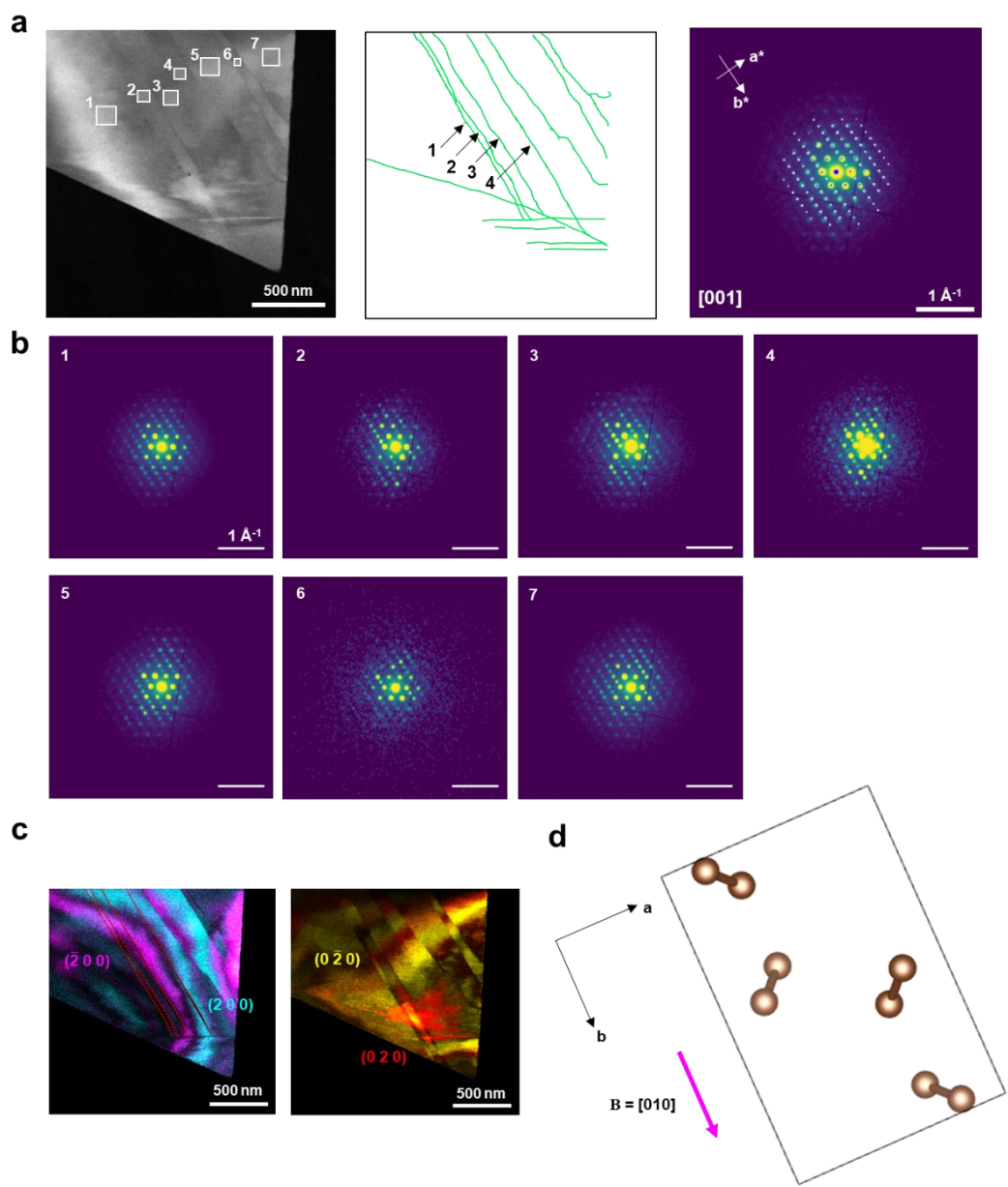

**Supplementary Figure 16. Analysis of dislocation network in *n*-hentriacontance sample. a,** ADF image showing dislocation lines and the shift of a bend contour on crossing a dislocation. Several dislocations are presented by green lines, showing the dense network across the film. Corresponding diffraction pattern (with on-zone simulated pattern overlaid in white) showing the paraffin film aligns on the [001] zone axis. **b,** Electron diffraction patterns taken at different areas between the dislocation lines showing the same in-plane orientation suggesting that these areas are not from different domains and that the lines are true dislocations. **c,** VDF images of the bend contour pairs at different **g.B** conditions. VDF image of the bend contour pairs showing the continuous bend contours on crossing dislocation-1, -2, -3, and -4 at $g_{200}$ and $g_{\bar{2}00}$ and the break in bend contours at $g_{020}$ and $g_{0\bar{2}0}$. The Burgers vector for these dislocations follows **B** = [010]. **c,** Modelling of the **B** direction in paraffin unit cell viewing along [001] with adjusted in-plane rotation to match the orientation in the diffraction pattern from experimental data set.

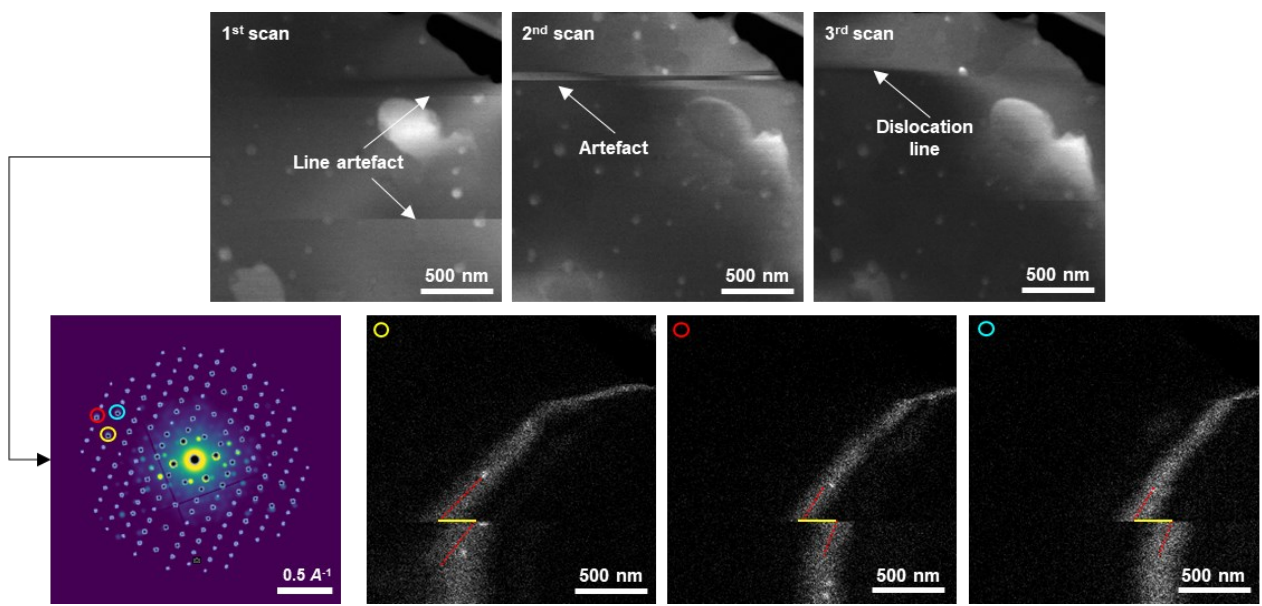

**Supplementary Figure 17. Artefacts arising from SED due to the scanning of the electron beam on the sample for data set taken at 200 kV.** Series of ADF images from repeated SED measurement on the same area of p-terphenyl film showing the appearance of line artefacts parallel to the scanning direction of the electron beam which resemble dislocation lines. The line artefacts can be distinguished from true dislocations by their disappearance under repeated SED measurement. The displacement of the bend contours due to the line artefacts showing no differences in the magnitude at different diffraction vectors $\mathbf{g}_{hkl}$ that are not parallel, as marked in the diffraction pattern with on-zone simulated pattern overlaid in white, and this finding can be used to also distinguish line artefacts and true dislocations. In addition, an abrupt contrast change at the middle area between the 1st scan, 2nd scan, and 3rd scan indicates the movement of the bend contour, i.e. sample reorientation, at each scan.

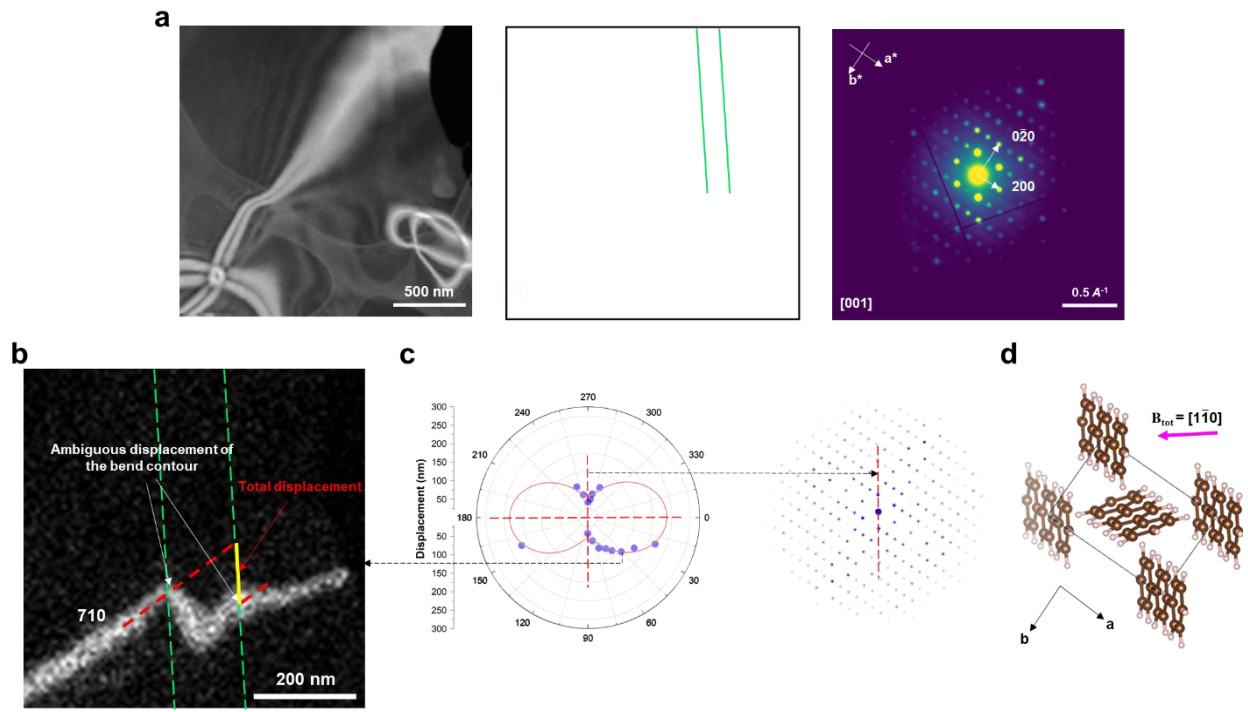

**Supplementary Figure 18. Adjacent, parallel dislocation lines in p-terphenyl.** This dataset was taken at 200 kV. **a,** ADF image of the p-terphenyl film showing several pole bend contours within the field of view. One of the bright bend contour intersects with two parallel dislocation lines as presented by the green lines (middle panel). The corresponding diffraction pattern for the whole field of view is shown on the right, indicating that the film is oriented along the [001] zone axis. **b**, VDF image of the $g_{710}$ bend contour highlighting the limited separation and resulting challenge for reliable identification of bend contour displacements for each individual dislocation line. The total displacement of the bend contour on crossing both dislocations was measured to examine the information on the combined pair of dislocations. **c**, A polar plot showing total displacement of the bend contours on crossing both dislocations via azimuthal angle φ (with on-zone simulated pattern added next to the plot). The fitting function (equation 1) can be modified with an additional parameter $D$: $f(\varphi) = A \arctan(B \cos^2(\varphi - C)) + D$ to capture the observed response. The resulting plot and fit does not reach zero, possibly due to the contribution (sum) of two dislocations with different Burgers vectors or due to deviations from a linear and elastic response between the two dislocations. Estimation of quasi-invisibility criterion condition using the local minimum of the constructed polar plot and the simulated diffraction pattern showing the diffraction vector $g_{110}$ and $g_{\bar{1}\bar{1}0}$. One possible explanation is that the Burgers vectors of the two dislocations sum to $B_{tot} = [1\bar{1}0]$. **d,** The $B_{tot}$ direction in p-terphenyl unit cell viewing along [001] direction with adjusted in-plane rotation to match the orientation of the diffraction pattern.

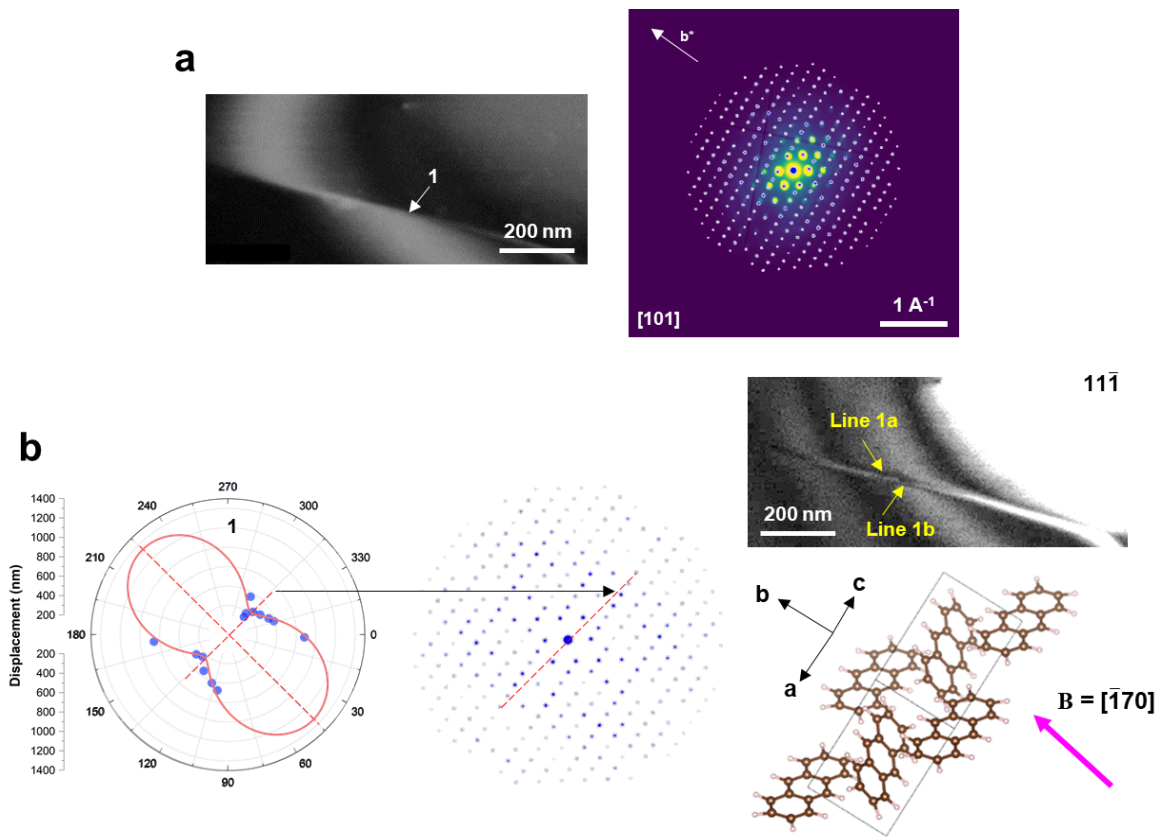

**Supplementary Figure 19. Adjacent, parallel dislocation lines in anthracene. a,** ADF image showing the shifting of a bright bend contour on crossing dislocation-1 and the corresponding diffraction pattern (with on-zone simulated pattern overlaid in white) indicating the film near the [101] zone axis. **b,** A polar plot of the bend contour displacement as a function of azimuthal angle φ constructed for dislocation-1 with on-zone simulated pattern added next to the plot. Fitting was carried out using a modified form of equation 1 with an additional fitting parameter $D$ (see also Supplementary Figure 18). (Right) VDF image constructed from $g_{11\bar{1}}$ shows two dislocation lines in very close proximity with near-parallel dislocation lines. The polar plot shows a quasi-invisibility criterion condition at ~ $g_{\bar{7}1\bar{7}}$ and $g_{71\bar{7}}$ for the combined dislocations. The extracted Burgers vector follows $\mathbf{B}_{tot} = [u_B, v_B, w_B]$ with $u_B$ = -0.11, $v_B$ = 1.00, $w_B$ = 0.03 which is approximately along $[u_B v_B 0]$ = $[\bar{1}70]$ (2.6° from perpendicular to [101] and 2.6° from the extracted **B**, 1.5 standard errors for a standard error of 0.03 radians). Modelling of the $\mathbf{B}_{tot}$ direction in anthracene unit cell viewing along [101] with adjusted in-plane rotation to match the orientation of the diffraction pattern from experimental dataset.

**Supplementary Note 4: Measurements of beam damage**

Critical fluence values are determined generally by examining the decay of a signal (e.g. diffraction intensity, image Fourier transform peak, spectroscopic signal) as a function of cumulative electron fluence. For a series of diffraction data acquired with constant exposure per unit time, the cumulative electron fluence can be calculated by multiplying the time that the samples had been exposed to the electron beam (in seconds) by the electron beam flux ($J$):

$$F(e^-Å^{-2}) = J \times (t_0 + t) \tag{S-16}$$

where $t_0$ is the time between the initial exposure of the analysed area to the electron beam and the time taken to record the image and the first diffraction pattern and $t$ is the subsequent acquisition time of the following diffraction patterns. In this work, electron diffraction patterns were analysed by the custom-made Python scripts using functions from HyperSpy package (1.6.5)[5]. The spot intensity ($I$) of the selected diffraction spot was extracted from a series of diffraction pattern acquisitions to construct a time series, or equivalently, a cumulative fluence series. The damage of the samples is assumed to occur as an exponential decay of the spot intensity with decay rate $\tau$ as a function of cumulative electron fluence[6]:

$$I = I_0 \exp(-\tau F(t)) \tag{S-17}$$

Rearrangement to normalize by the initial intensity ($I/I_0$) for the particular selected spot establishes a linear relationship between $ln(I/I_0)$ and the cumulative fluence. Fitting this linear response enables determination of the gradient, with the critical fluence (CF) defined as $1/\tau$.

**References**

1. Jones, W. & Williams, J. O. Real space crystallography and defects in molecular crystals. *J Mater Sci* **10**, 379–386 (1975).

2. Newville, M. *et al.* lmfit/lmfit-py: 1.2.1. (2023) doi:10.5281/zenodo.7887568.

3. Spiecker, E. & Jäger, W. Burgers vector analysis of large area misfit dislocation arrays from bend contour contrast in transmission electron microscope images. *J. Phys.: Condens. Matter* **14**, 12767–12776 (2002).

4. Cherns, D. & Preston, A. R. Convergent beam diffraction studies of interfaces, defects, and multilayers. *Journal of Electron Microscopy Technique* **13**, 111–122 (1989).

5. Peña, F. de la *et al.* hyperspy/hyperspy: Release v1.6.5. (2021) doi:10.5281/zenodo.5608741.

6. Ilett, M. *et al.* Analysis of complex, beam-sensitive materials by transmission electron microscopy and associated techniques. *Phil. Trans. Roy. Soc. A* **378**, 20190601 (2020).